\documentclass[twocolumn]{aastex61}
\usepackage{amsmath}
\submitjournal{ApJ}
\accepted{September 20, 2018}

\shorttitle{SDP.11}
\shortauthors{Lamarche et al.}

\begin{document}

\title{Resolving Star Formation on Sub-Kiloparsec Scales in the High-Redshift Galaxy SDP.11 Using Gravitational Lensing\footnote{\emph{Herschel} is an ESA space observatory with science instruments provided by European-led Principal Investigator consortia and with important participation from NASA.}}

\correspondingauthor{Cody Lamarche}
\email{cjl272@cornell.edu}

\author{C. Lamarche}
\affil{Department of Astronomy, Cornell University, Ithaca, NY 14853.}

\author{A. Verma}
\affil{Sub-department of Astrophysics, University of Oxford, Oxford OX1 3RH.}

\author{A.Vishwas}
\affil{Department of Astronomy, Cornell University, Ithaca, NY 14853.}

\author{G. J. Stacey}
\affil{Department of Astronomy, Cornell University, Ithaca, NY 14853.}

\author{D. Brisbin}
\affil{N$\acute{\rm{u}}$cleo de Astronom$\acute{\rm{i}}$a, Facultad de Ingenier$\acute{\rm{i}}$a y Ciencias, Universidad Diego Portales, Av. Ej$\acute{\rm{e}}$rcito 441, 8370191 Santiago, Chile.}

\author{C. Ferkinhoff}
\affil{Department of Physics, Winona State University, Winona, MN, 55987.}

\author{T. Nikola}
\affil{Cornell Center for Astrophysics and Planetary Science, Cornell University, Ithaca, NY 14853.}

\author{S. J. U. Higdon}
\affil{Department of Physics, Georgia Southern University, Statesboro, GA 30460.}

\author{J. Higdon}
\affil{Department of Physics, Georgia Southern University, Statesboro, GA 30460.}

\author{M. Tecza}
\affil{Sub-department of Astrophysics, University of Oxford, Oxford OX1 3RH.}

\begin{abstract}

We investigate the properties of the interstellar medium, star formation, and the current-day stellar population in the strongly-lensed star-forming galaxy H-ATLAS J091043.1-000321 (SDP.11), at z = 1.7830, using new Herschel and ALMA observations of far-infrared fine-structure lines of carbon, oxygen and nitrogen. We report detections of the [O\,{\sc iii}] 52 $\micron$, [N\,{\sc iii}] 57 $\micron$, and [O\,{\sc i}] 63 $\micron$ lines from Herschel/PACS, and present high-resolution imaging of the [C\,{\sc ii}] 158 $\micron$ line, and underlying continuum, using ALMA. We resolve the [C\,{\sc ii}] line emission into two spatially-offset Einstein rings, tracing the red- and blue-velocity components of the line, in the ALMA/Band-9 observations at 0$\farcs$2 resolution. The values seen in the [C\,{\sc ii}]/FIR ratio map, as low as $\sim$ 0.02\% at the peak of the dust continuum, are similar to those of local ULIRGs, suggesting an intense starburst in this source. This is consistent with the high intrinsic FIR luminosity ($\sim$ 3 $\times$ 10$^{12}$ L$_\odot$), $\sim$ 16 Myr gas depletion timescale, and $\lesssim$ 8 Myr timescale since the last starburst episode, estimated from the hardness of the UV radiation field. By applying gravitational lensing models to the visibilities in the uv-plane, we find that the lensing magnification factor varies by a factor of two across SDP.11, affecting the observed line profiles. After correcting for the effects of differential lensing, a symmetric line profile is recovered, suggesting that the starburst present here may not be the result of a major merger, as is the case for local ULIRGs, but instead could be powered by star-formation activity spread across a 3-5 kpc rotating disk.

\end{abstract}


\section{Introduction} \label{sec:intro}
One of the key goals of modern astrophysics is to understand the processes which govern star formation and galaxy assembly in the early Universe. The epoch of peak star-formation rate density, which occurred between 8-11 billion years ago (1\,\textless\,z\,\textless\,3), is of particular interest to understanding the assembly of present-day massive elliptical galaxies.  Within this epoch, most of the star formation is obscured by a heavy veil of dust \citep[e.g.,][]{Casey2014}. This dust absorbs stellar ultraviolet (UV) radiation and re-emits it thermally in the far-infrared (FIR). The best probes of the physical conditions within these dusty star-forming galaxies (DSFGs) therefore lie in the far-IR and include several bright fine-structure lines that emanate from astrophysically abundant species such as carbon, nitrogen, and oxygen. These FIR lines arise from energy levels in the ground state term whose degeneracy is broken by spin-orbit coupling. As such, they are easily excited at modest (few hundred K) gas temperatures. These FIR lines trace the physical conditions of the gas, often being important sources of gas cooling, such that they reveal the properties of the sources which heat the gas. 

For example, the [C\,{\sc ii}] 158 $\micron$ line largely arises from warm, dense, photodissociation regions (PDRs) on the surfaces of molecular clouds and the [C\,{\sc ii}] to FIR continuum luminosity ratio is a sensitive measure of the far-UV (6 to 13.6 eV) radiation field intensity, G$_0$, and hence star formation intensity \citep[e.g.,][]{Wolfire1990}. The [O\,{\sc i}] 63 $\micron$ line arises from deeper within PDRs and, together with the [C\,{\sc ii}] line and FIR continuum, constrains the PDR gas density and further refines the estimate of G$_0$.

The [O\,{\sc iii}] 88 and 52 $\micron$, [N\,{\sc ii}] 122 and 205 $\micron$, and [N\,{\sc iii}] 57 $\micron$ lines originate in ionized gas regions. The typical temperature of these H\,{\sc ii} regions is $\sim$ 8,000\,K, and hence the populations in the FIR line-emitting levels, which lie a few hundred K above ground, are primarily sensitive to the density of the medium. Thus, line ratios within a given ionic species yield H\,{\sc ii} region density. Hardness of the radiation field strongly affects the ionization equilibrium of metal ions, such that comparing the relative brightness of lines emitted from ions with significantly different ionization potentials allows us to constrain that property. For example, the [O\,{\sc iii}] 88 $\micron$ and the [N\,{\sc ii}] 122 $\micron$ lines have nearly identical critical densities (510\,cm$^{-3}$ and 310\,cm$^{-3}$, respectively, at 8,000\,K) but significantly different formation potentials (O$^{++}$: 35 eV and N$^{+}$: 14 eV). Hence, the [O\,{\sc iii}]/[N\,{\sc ii}] line ratio probes the hardness of the stellar radiation field, and thus the high-mass end of the current day stellar mass function \citep[e.g.,][]{Ferkinhoff2011}. The combination of these ionized gas lines provides tight constrains on the UV field hardness, and enables us to make estimates of the N/O abundance ratio \citep[e.g.,][]{Lester1987}. 

Each of these lines have been surveyed for galaxies in the local Universe \citep[e.g.,][]{Crawford1986, Stacey1991, Malhotra1997, Malhotra2001, Gracia-Carpio2011, Parkin2013, Cormier2015, Hughes2015, Herrera-Camus2016, Diaz-Santos2017} and [C\,{\sc ii}] and [O\,{\sc i}] surveys of high-redshift galaxies have appeared \citep[e.g.,][]{Stacey2010, Brisbin2015, Gullberg2015, Coppin2012}. In this paper, we study the ISM and star forming conditions within H-ATLAS J091043.1-000321 (hereafter SDP.11), a strongly-lensed, Ultra Luminous Infrared Galaxy- (ULIRG-) like source, at z = 1.7830, using multiple far-IR fine-structure lines of carbon, nitrogen and oxygen.

SDP.11 was first identified as a potential high-redshift, gravitationally-lensed, source in the Science Demonstration Phase of the Herschel Astrophysical Terahertz Large Area Survey \citep[H-ATLAS,][]{Eales2010} due to its large observed 500 $\micron$ flux \citep{Negrello2010}. Follow-up observations conducted with Z-Spec on the Caltech Submillimeter Observatory (CSO) detected several mid-J CO lines, consistent with a redshift of 1.786 $\pm$ 0.005 \citep{Lupu2012}.

\cite{Bussmann2013} presented Sub-Millimeter Array (SMA) observations of the thermal dust continuum at 880 $\micron$ (observed-frame) that revealed two images of the lensed galaxy separated by $\sim$ 2$\farcs$2 on the sky. \cite{Negrello2014} identified an elliptical Einstein ring in near-infrared images taken with the Wide-Field Camera-3 on-board the Hubble Space Telescope (HST/WFC3). Using these data, the magnification factor due to gravitational lensing was estimated to be $\sim$ 8 based on stellar emission \citep{Dye2014} and $\sim$ 11 based on thermal dust emission \citep{Bussmann2013}. The lensing galaxy was found to coincide with the position of a known optical source at z = 0.792.

\cite{Ferkinhoff2014} first reported the detection of the [C\,{\sc ii}] 158 $\micron$ line with the second-generation redshift (Z) and Early Universe Spectrometer (ZEUS-2) on the Atacama Pathfinder Experiment (APEX) telescope. Combined with an estimate of the FIR luminosity and a preliminary estimate of the [O\,{\sc i}] 63 $\micron$ line flux from Herschel/PACS, they constrained the source-averaged physical conditions of the photo-dissociation regions in SDP.11. They suggest that SDP.11 hosts an intense and dense starburst (G$_0$ $\sim$ 20,000 Habing units, n $\sim$ 2,300 cm$^{-3}$), as evidenced by the low L$_{[C\,\textrm{\sc ii}]}$/L$_{FIR}$ ratio, (1.0 $\pm$ 0.3) $\times$ 10$^{-3}$, analogous to that of local ULIRGs, possibly driven by a merger.

Here we present follow-up observations of the [C\,{\sc ii}] 158 $\micron$ line, conducted with the Atacama Large Millimeter/submillimeter Array (ALMA), at a spatial resolution of 0$\farcs$2. We also present strong detections of a suite of far-IR fine-structure lines, arising from both neutral and ionized gas, observed with the PACS spectrometer onboard the Herschel Space Observatory, as well as multi-band radio continuum observations conducted with NSF's Karl G. Jansky Very Large Array (VLA). We combine these datasets with Herschel/SPIRE observations to constrain the time since the last starburst, estimate the gas-phase [N/O] abundance ratio, perform lens modeling of SDP.11 to recover the intrinsic (unlensed) properties of the source, and examine the variation in the [C\,{\sc ii}]/FIR ratio on 500 pc spatial scales.

We assume a flat $\Lambda$CDM cosmology, with $\Omega_M$ = 0.27, $\Omega_\Lambda$ = 0.73, and H$_0$ = 71\,km\,s$^{-1}$\,Mpc$^{-1}$, throughout this paper \citep{Spergel2003}, such that 1" = 8.54 kpc, D$_A$ = 1.76 Gpc, and D$_L$ = 13.65 Gpc.

\section{Observations and Data Reduction} \label{sec:observations}

\subsection{ALMA} \label{subsec:ALMA}

The [C\,{\sc ii}] 158 $\micron$ line was observed in SDP.11 using the Atacama Large Millimeter/submillimeter Array (ALMA)\footnote{The National Radio Astronomy Observatory is a facility of the National Science Foundation operated under cooperative agreement by Associated Universities, Inc.} Band 9 receivers. The observations were conducted on November 16, 2016, with the array in the C40-4 configuration, using 42 antennas, with baselines ranging from 15 to 920 m, under excellent weather conditions, with a precipitable water vapor (PWV) measurement of 0.28 mm. Observing at 683 GHz in this array configuration, the interferometer is sensitive to a maximum recoverable scale of $\sim$ 1$\farcs$3. The total on-source integration time for these observations was 12.6 minutes.

For these observations, J0854+2006, J0522-3627, and J0909+0121 were used as the bandpass, flux, and phase calibrators, respectively. The data were reduced, imaged, and cleaned using the Common Astronomy Software Application (CASA)\footnote{https://casa.nrao.edu/}, version 4.7.2.

The [C\,{\sc ii}] data were imaged using 50\,km\,s$^{-1}$ spectral channels, and natural weighting, achieving a synthesized beam of size 0$\farcs$20 x 0$\farcs$16. The RMS sensitivity is $\sim$\,4.5\,mJy\,beam$^{-1}$ in each 50\,km\,s$^{-1}$ channel.

A continuum image was created by combining all non-line spectral channels present in the measurement set, for a total continuum bandwidth of $\sim$ 6.5 GHz, which, when imaged similarly to the [C\,{\sc ii}] line, yields a beam size of 0$\farcs$20 x 0$\farcs$15 and an RMS sensitivity of 0.72\,mJy\,beam$^{-1}$.

\subsection{Herschel/PACS} \label{subsec:PACS}

The [O\,{\sc iv}] 26 $\micron$, [S\,{\sc iii}] 33 $\micron$, [O\,{\sc iii}] 52 $\micron$, [N\,{\sc iii}] 57 $\micron$, and [O\,{\sc i}] 63 $\micron$ fine-structure lines were all observed in SDP.11 using the Photodetector Array Camera and Spectrometer (PACS) \citep{Poglitsch2010} onboard the Herschel Space Observatory \citep{Pilbratt2010} (OBS ID's: 1342231291, 1342231292, 1342231293, and 1342231294).  All of these observations were conducted on October 20, 2011, using the instrument in the RangeSpec mode, with a duration of $\sim$ 30 - 90 minutes per observation, and central pointing coordinates of (9$^h$10$^m$43$^s$.1, -00$^o$03'24".0). The velocity resolution of these observations is $\sim$ 110 km s$^{-1}$, with a typical 1$\sigma$ statistical noise of $\sim$ 35 - 85 mJy per velocity bin.

The raw data were processed using the Herschel Interactive Pipeline Environment (HIPE) \citep{Ott2010} version 15.0.1. A point-source correction was applied to the spectrum extracted from the central 9$\farcs$4 x 9$\farcs$4 spatial pixel (spaxel), since the diameter of the Einstein ring is only $\sim$ 2$\farcs$2.

\subsection{Herschel/SPIRE} \label{subsec:SPIRE}

The [O\,{\sc iii}] 88 $\micron$, [N\,{\sc ii}] 122 $\micron$, and [C\,{\sc ii}] 158 $\micron$ lines were all observed in SDP.11 using the Spectral and Photometric Imaging Receiver (SPIRE) \citep{Griffin2010} onboard the Herschel Space Observatory and were first presented in \cite{Zhang2018}.

The raw data were processed using the Herschel Interactive Pipeline Environment (HIPE) \citep{Ott2010} version 15.0.1, with SPIRE calibration version 14.3. The baselines of the resulting spectra were corrected for instrumental effects using the off-source detectors. The continuum was fitted with a second-order polynomial and absolute flux calibration was verified by comparing synthetic photometry generated from the spectra, using the HIPE script ``spireSynthPhotometry", to SPIRE photometer maps of SDP.11.

\section{Results and Discussion} \label{sec:ReasultsAndDiscussion}

\subsection{Line and Continuum Fluxes} \label{subsec:fluxes}

The [C\,{\sc ii}] 158 $\micron$ line is strongly detected in the ALMA observations, emanating from a nearly complete, elliptical, Einstein ring with a diameter of $\sim$ 2$\farcs$2 and an axial ratio of $\sim$ 0.8 (see Figure 1). The line has two clearly defined velocity components, separated by $\sim$ 310\,km\,s$^{-1}$, which appear as two spatially-offset Einstein rings on the sky. Creating a moment-zero, primary-beam corrected, map by collapsing the spectral cube along the velocity axis, and summing the flux from pixels detected at $>$ 3$\sigma$ in either the red or blue component of the line in the flat-noise maps, we calculate a source-integrated [C\,{\sc ii}] 158 $\micron$ flux of 260 $\pm$ 9\,Jy\,km\,s$^{-1}$, or equivalently (5.9 $\pm$ 0.2)\,$\times$\,10$^{-18}$\,W\,m$^{-2}$, where the uncertainties are estimated by propagating the RMS error per beam over the line-emitting region of the source. This method of calculating the flux by creating a mask using the flat-noise map and then applying that mask to the primary-beam corrected map is employed because the pointing of our ALMA observations was offset from the center of SDP.11 by $\sim$ 2", such that the noise around the source is non-uniformly amplified by the necessary primary beam correction.

In contrast to the stellar emission seen in HST/WF3 near-IR images, and the [C\,{\sc ii}] line emission, the rest-frame 158 $\micron$ continuum, measured from the ALMA observations, emanates prominently from two locations along the Einstein ring of SDP.11, one in the north and one in the south, which lie neatly between the red and blue Einstein rings seen in the [C\,{\sc ii}] line emission (see Figure 1). Again, summing the flux density in the primary-beam corrected map from pixels detected at $>$ 3$\sigma$ in the flat-noise map, we obtain a source-integrated specific flux of 189 $\pm$ 4 mJy at 158 $\mu$m rest-frame, where the uncertainty is estimated by propagating the RMS error per beam over the continuum-emitting region of the source. We note that this 3$\sigma$ hard cut should be considered a lower-limit on the total flux density, since it does not consider extended flux which may be present at lower significance. We expect that extended flux should be present beyond what we observe, given that SPIRE photometry measures 232 $\pm$ 8 mJy at 500 $\mu$m \citep[observed-frame;][]{Bussmann2013}, the closest photometric point to our continuum measurement at 440 $\mu$m (observed-frame).

\begin{figure*}[h]
\gridline{\fig{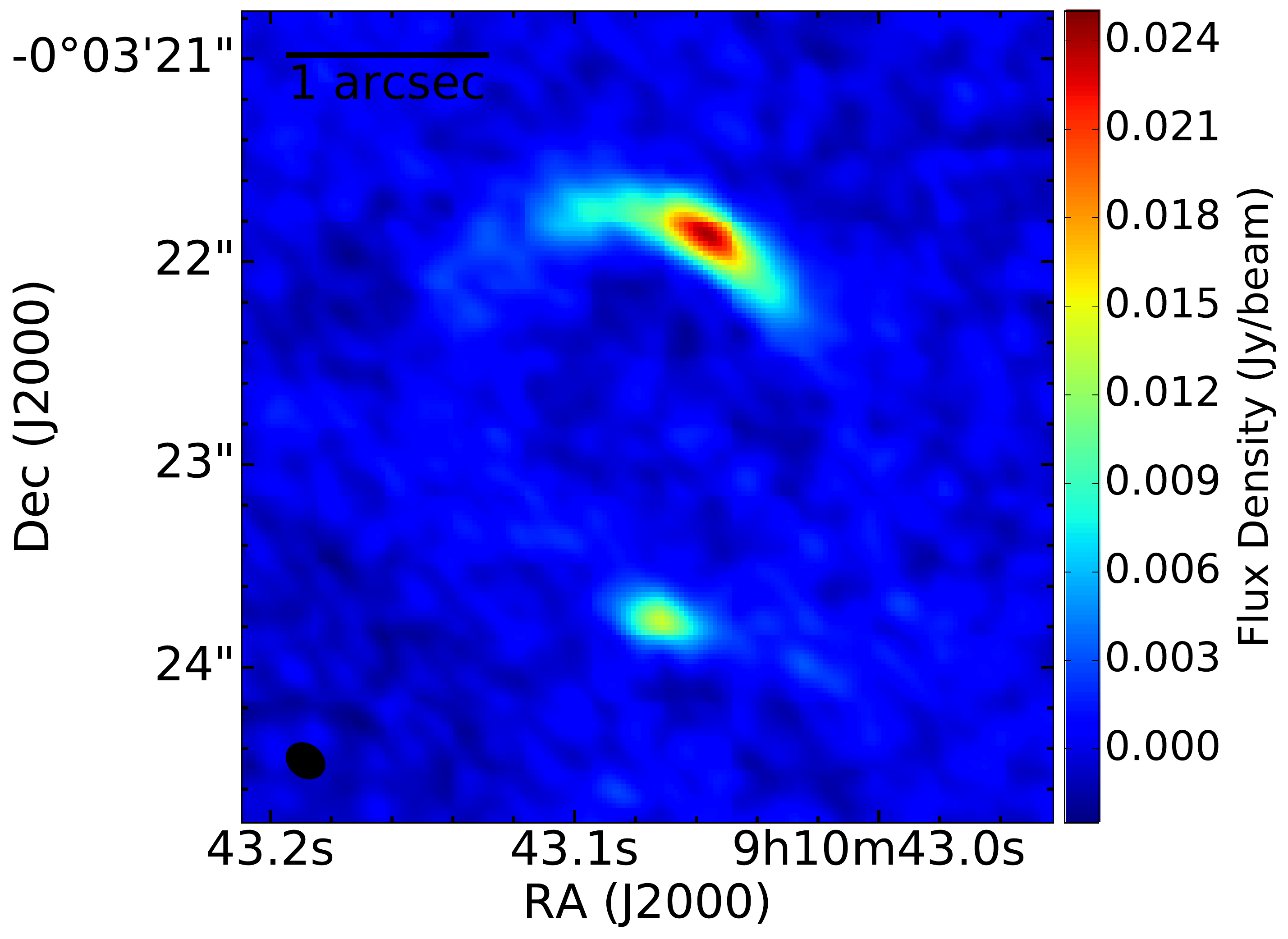}{0.5\textwidth}{(a)}
          \fig{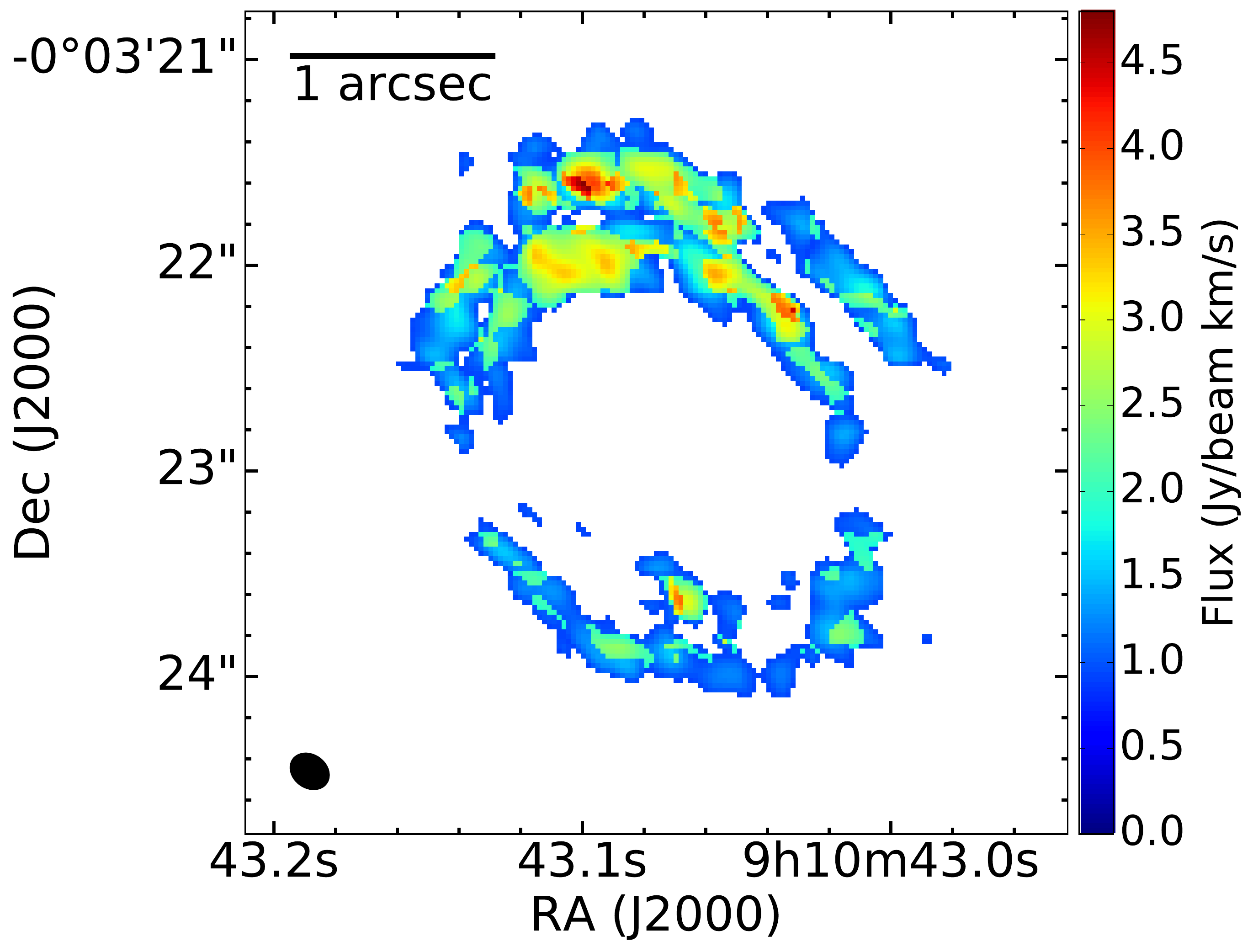}{0.48\textwidth}{(b)}
          }
\gridline{\fig{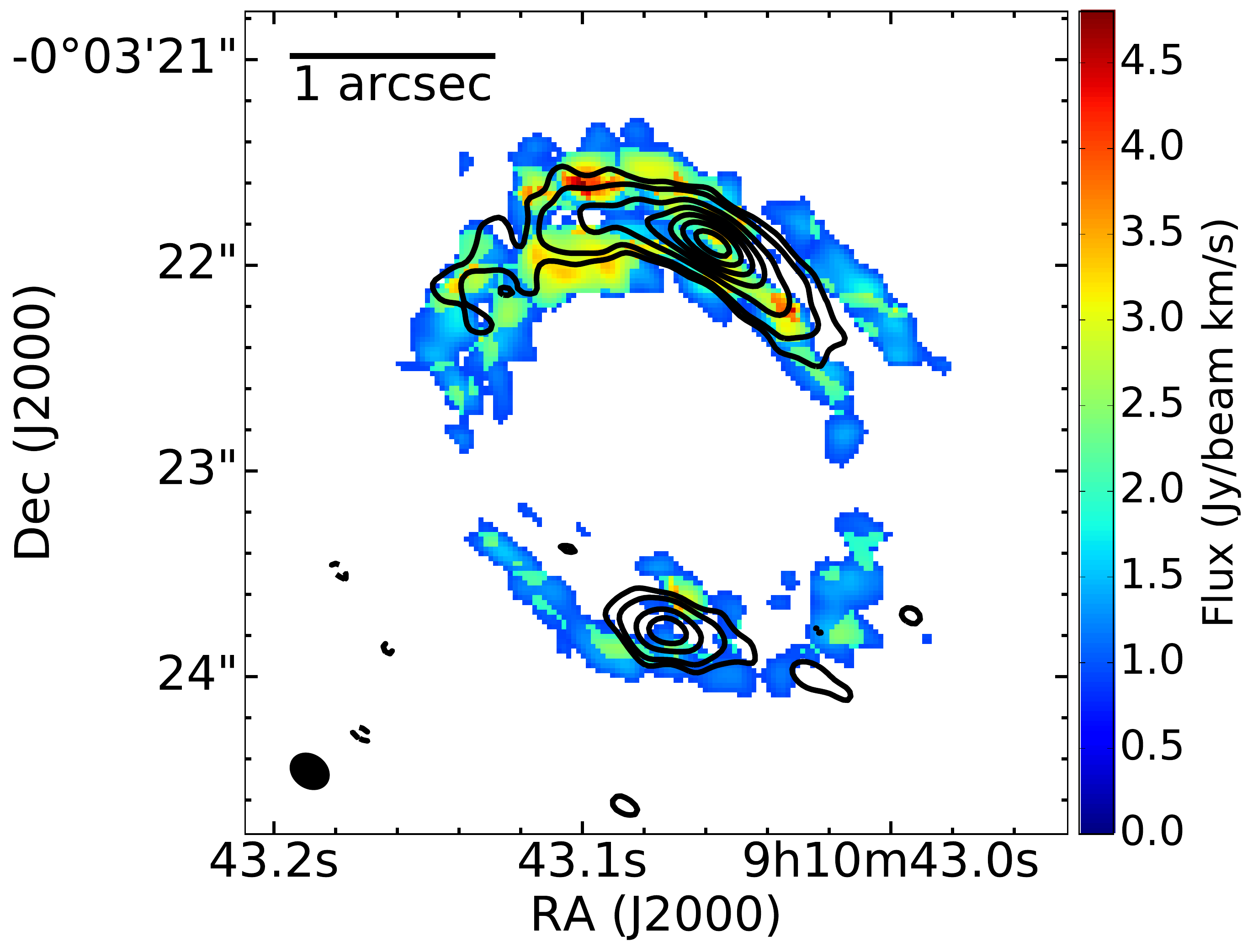}{0.48\textwidth}{(c)}
          \fig{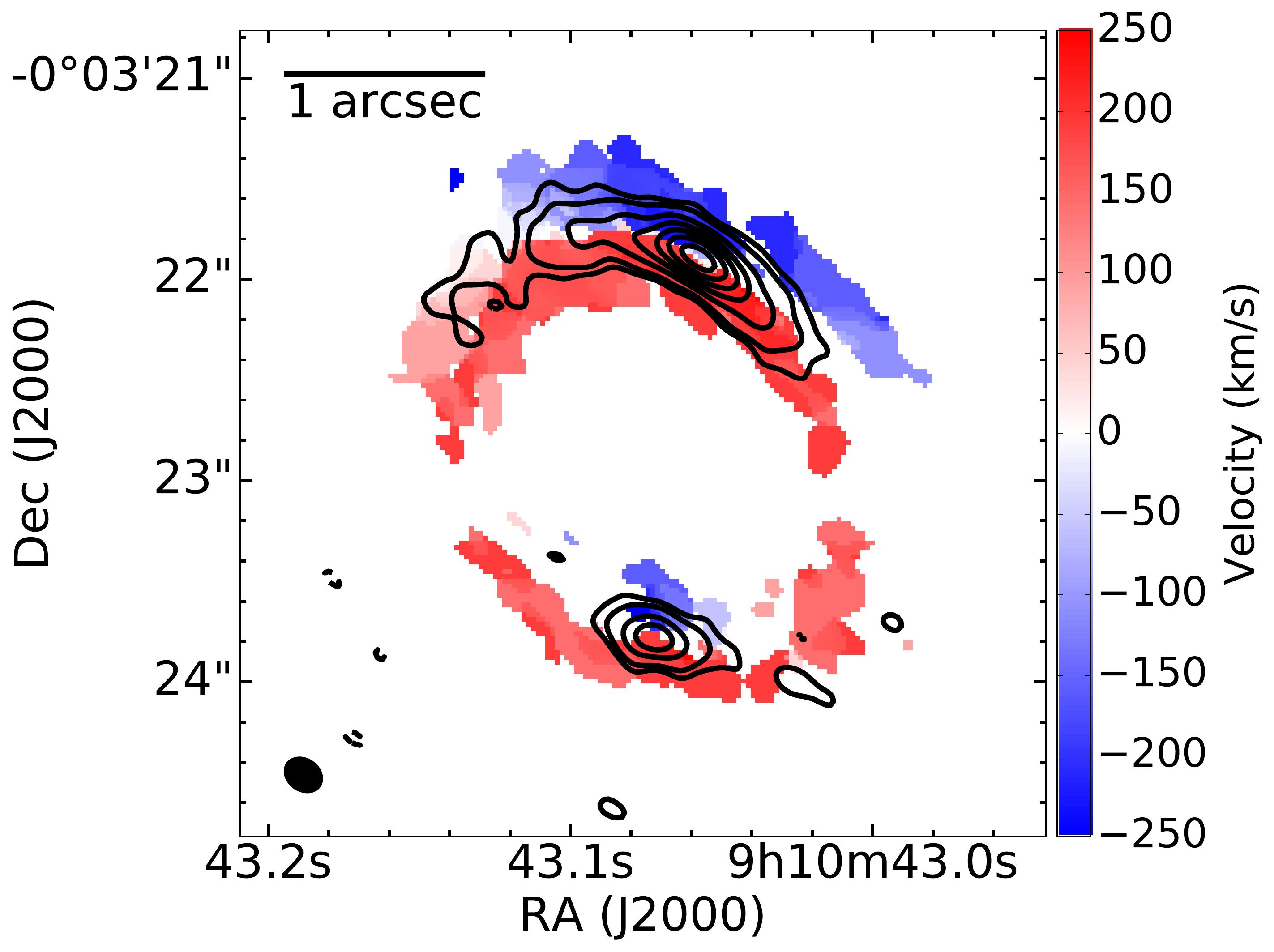}{0.5\textwidth}{(d)}
          }
\caption{(a) SDP.11 rest-frame 158 $\mu$m continuum color map, at 0$\farcs$2 resolution. The dust continuum emanates prominently from two locations along the Einstein ring. (b) SDP.11 [C\,{\sc ii}] 158 $\mu$m moment-zero color map, at 0$\farcs$2 resolution. The moment-zero map was created using a threshold of 4$\sigma$ per velocity channel. A nearly complete Einstein ring is visible in the line emission which is clearly resolved in the radial direction. (c) SDP.11 [C\,{\sc ii}] 158 $\mu$m moment-zero color map, with rest-frame 158 $\mu$m continuum contours superimposed (-5, -3, 3, 5, 10, 15, 20, 25, 30$\sigma$, negative contours dashed), at 0$\farcs$2 resolution. (d) SDP.11 [C\,{\sc ii}] 158 $\mu$m moment-one map, created using a threshold of 4$\sigma$ per velocity channel, with the same rest-frame 158 $\mu$m continuum contours as in (c) over-plotted. The two velocity components of the [C\,{\sc ii}] line are spatially offset on the sky, with the continuum emission centered between them. \label{fig:ALMAObs}}
\end{figure*}

The ALMA observations of the [C\,{\sc ii}] 158 $\micron$ line in SDP.11 clearly resolve the line into two velocity components (see Figure 2). We fit the source-integrated spectrum with a dual Gaussian line profile, one for each velocity component. We suggest a refined redshift for the source of 1.7830 $\pm$ 0.0002 for the [C\,{\sc ii}] 158 $\micron$ line, calculated as the average redshift of the two line components, with the error taken from the uncertainty in the Gaussian fitting. This redshift is consistent with the value of 1.786 $\pm$ 0.005 reported in \cite{Lupu2012}, determined from Z-Spec observations of several mid-J CO lines. From this updated redshift, the [C\,{\sc ii}] line components are located at v = -155 $\pm$ 18 km\,s$^{-1}$ and 155 $\pm$ 5 km\,s$^{-1}$, and have integrated fluxes of 124 $\pm$ 21\,Jy\,km\,s$^{-1}$ and 185 $\pm$ 14\,Jy\,km\,s$^{-1}$, respectively. 

\begin{figure*}
\plotone{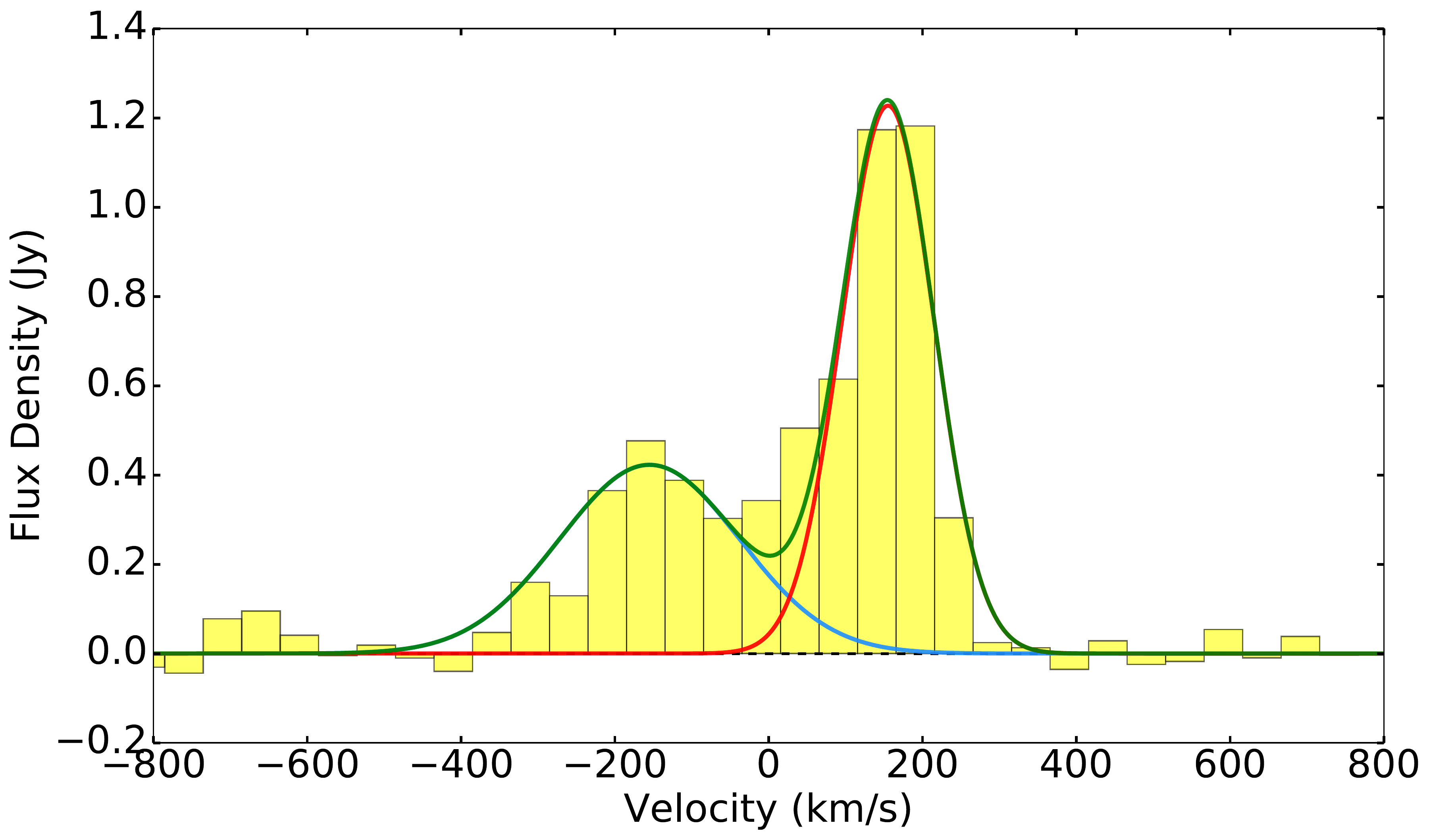}
\caption{Source-integrated ALMA [C\,{\sc ii}] 158 $\micron$ spectrum of SDP.11 (v = 0 corresponds to z = 1.7830). The line is clearly resolved into two velocity components, which we fit with two Gaussian line profiles (blue and red curves). The sum of the two components is shown in green. \label{fig:ALMASpectrum}}
\end{figure*}

We use the best-fit [C\,{\sc ii}] line velocities as priors for fitting the Herschel/PACS spectra, only allowing the central line velocity to vary by one PACS spectral bin in either direction, and varying the line widths and intensities. The resulting line fluxes are presented in Table 1, with the spectra and over-plotted best-fit line profiles in Figure 3.

\begin{figure}
\gridline{\fig{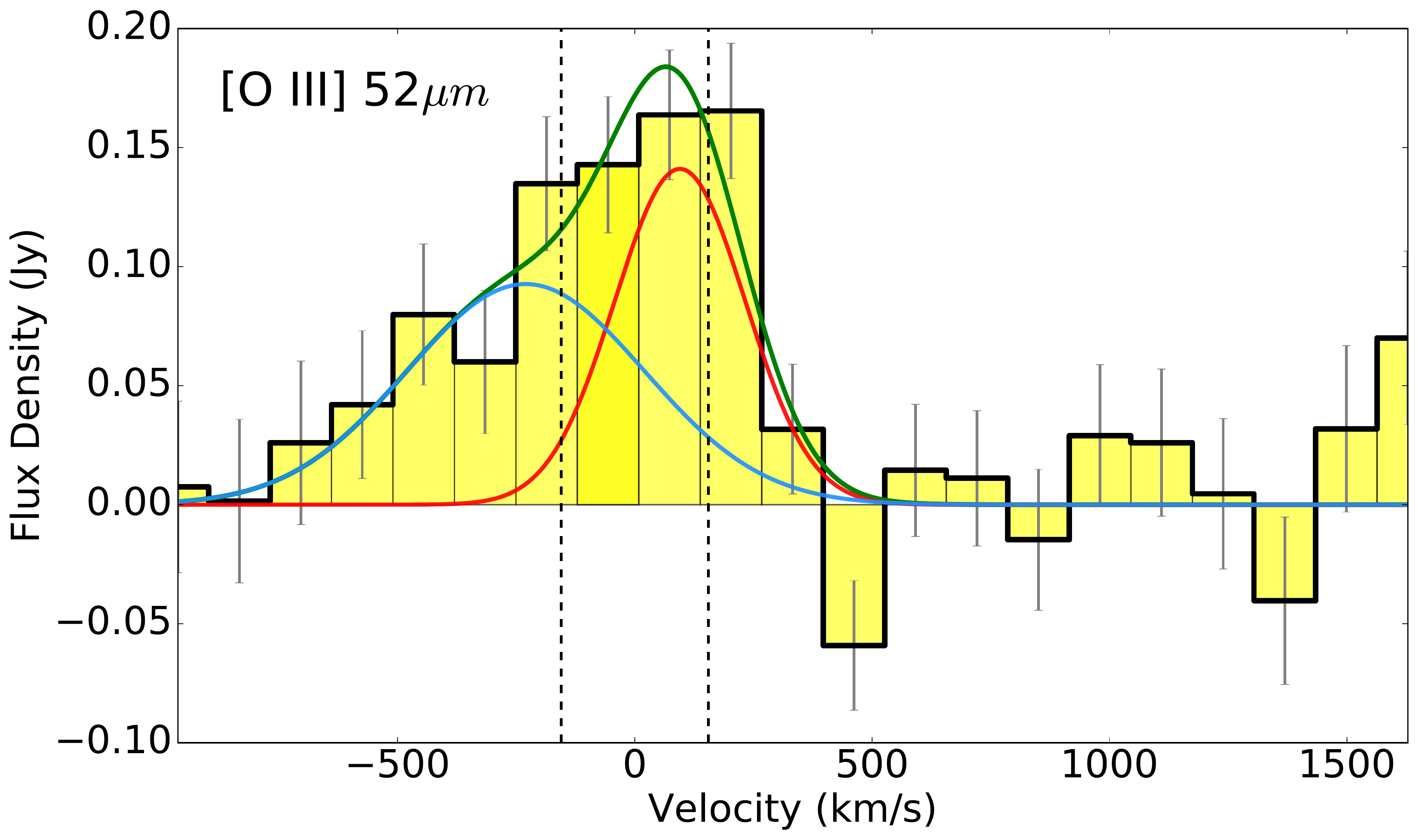}{0.48\textwidth}{(a)}
          }
\gridline{\fig{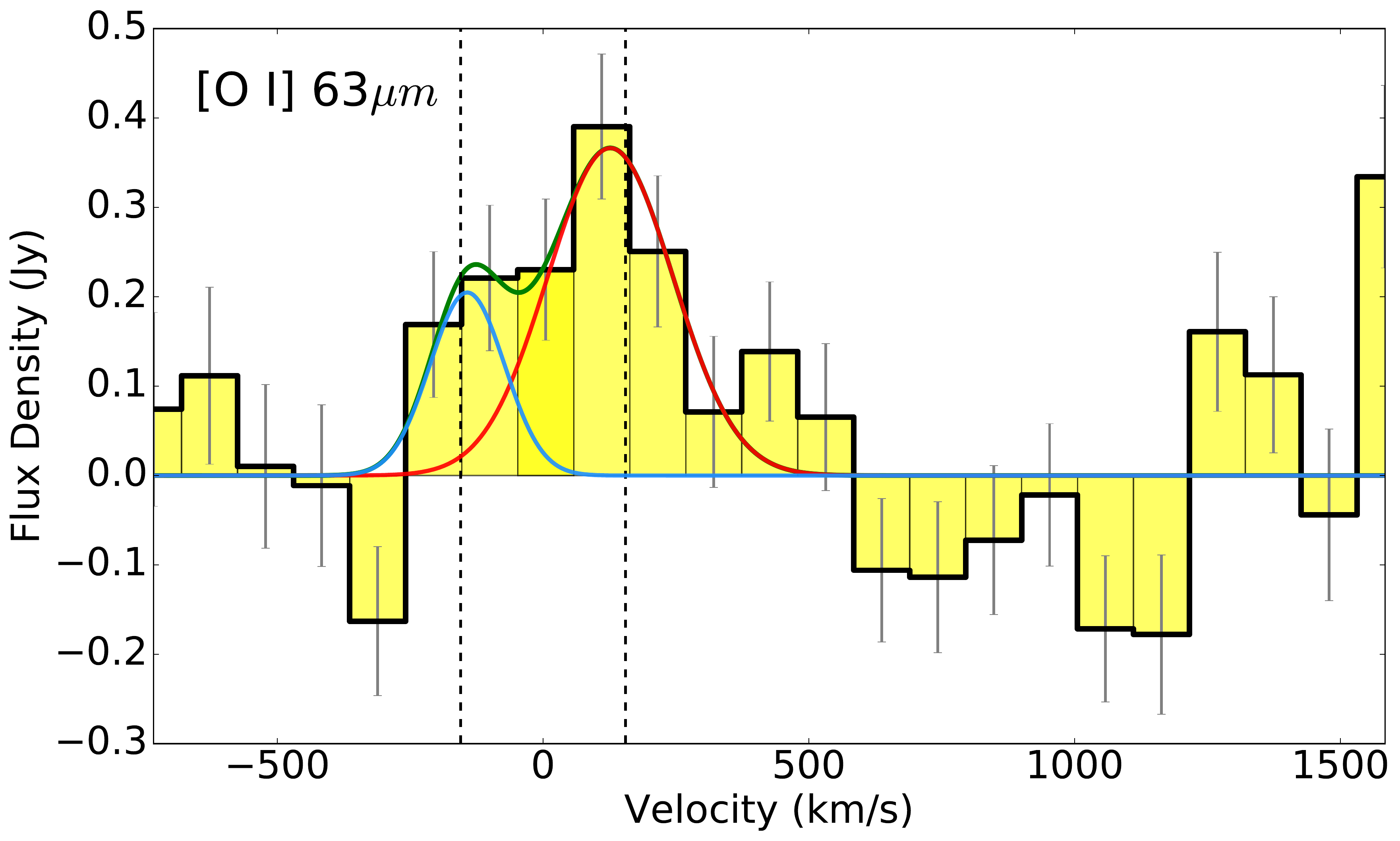}{0.48\textwidth}{(b)}
          }
\gridline{\fig{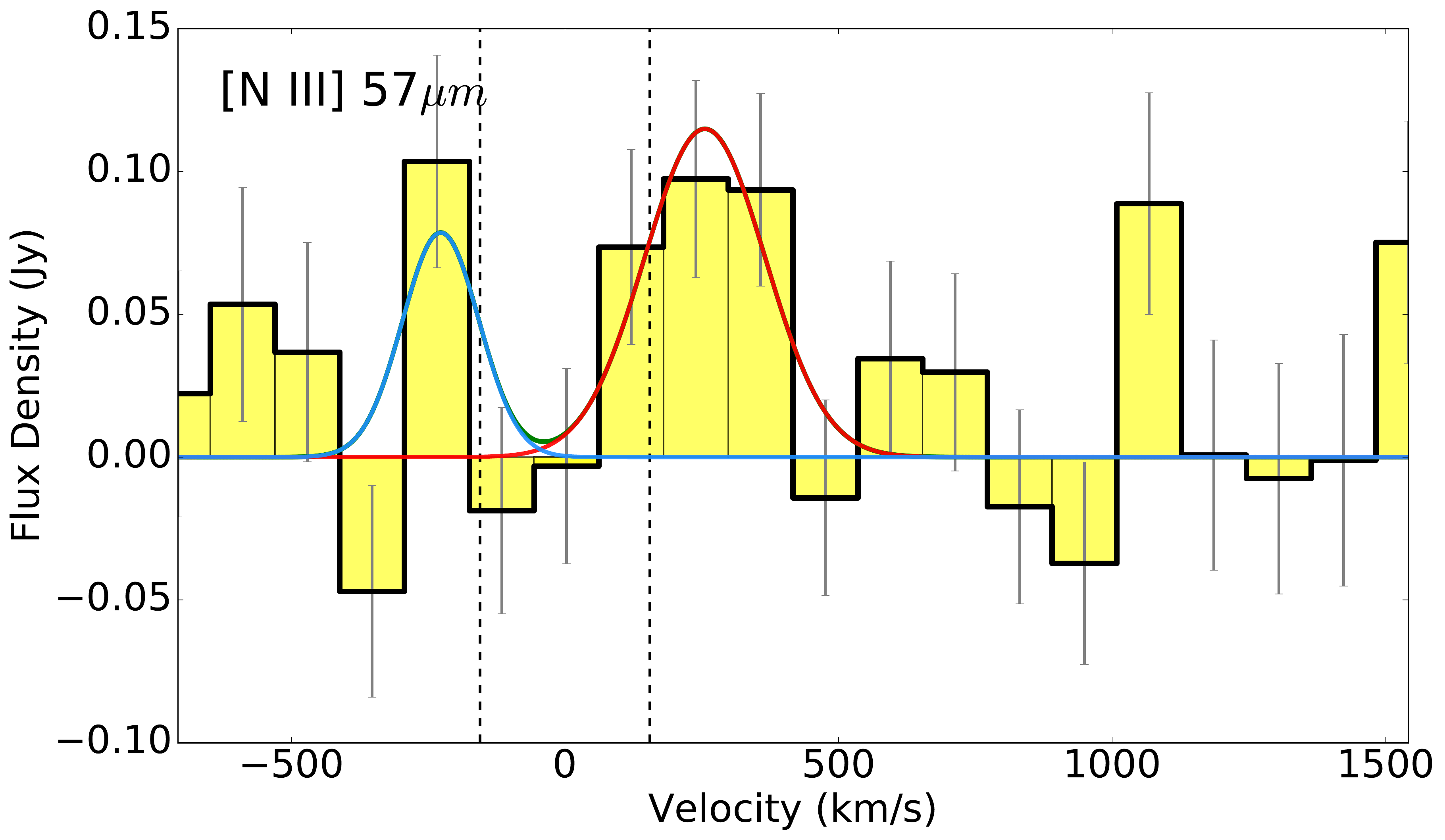}{0.48\textwidth}{(c)}
          }
\caption{Herschel/PACS spectra of the (a) [O\,{\sc iii}] 52 $\mu$m (130 km\,s$^{-1}$ bins), (b) [O\,{\sc i}] 63 $\mu$m (105 km\,s$^{-1}$ bins), and (c) [N\,{\sc iii}] 57 $\mu$m (118 km\,s$^{-1}$ bins) lines in SDP.11, plotted with 1$\sigma$ statistical error bars (v = 0 corresponds to z = 1.7830). We fit each of the lines with two Gaussian components (red and blue curves), with the sum of the two components shown in green. The positions of the two line components observed in the ALMA [C\,{\sc ii}] 158 $\micron$ spectrum are indicated by vertical dashed lines. The spectral lines observed with Herschel/PACS seem to have the same asymmetric line profile as does the [C\,{\sc ii}] 158 $\micron$ line. \label{fig:PACSObs}}
\end{figure}

The 3$\sigma$ limits for the non-detected lines observed with Herschel/PACS, also reported in Table 1, were determined by calculating the standard deviation of the baseline at the native spectral resolution of the instrument, at the wavelength of interest, and then binning up to an assumed line width of 500 km\,s$^{-1}$.

Fluxes for the lines observed with SPIRE were calculated using the built-in HIPE Spectrum Fitter. HIPE documentation recommends using a Sinc-Gauss model, which combines an intrinsic Gaussian line profile with the Sinc instrument response function, for fitting marginally-resolved spectral lines. We used the built-in HIPE Spectrum-fitter and simultaneously fit Sinc-Gauss models for the well detected [C\,{\sc ii}] and [O\,{\sc iii}] lines, using priors for the line position, width, and separation between the two components, from the ALMA [C\,{\sc ii}] spectrum, thus our reported fluxes differ slightly from \cite{Zhang2018}.

While the [C\,{\sc ii}] 158 $\micron$ flux which we measure with ALMA is consistent with our previous line flux measurement from APEX/ZEUS-2, (6.44 $\pm$ 0.42)\,$\times$\,10$^{-18}$ W\,m$^{-2}$ \citep{Ferkinhoff2014}, it is notably lower than the flux obtained from the SPIRE measurements, (12.9 $\pm$ 1.3)\,$\times$\,10$^{-18}$ W\,m$^{-2}$. The same flux discrepancy is found by \cite{Zhang2018}, between their reported APEX/SEPIA and SPIRE fluxes. This flux discrepancy could be partially due to the diameter of the Einstein ring, $\sim$ 2$\farcs$2, being larger than the maximum recoverable scale of the utilized ALMA array configuration in band 9, $\sim$ 1$\farcs$3, such that we are resolving out flux with the interferometer. However, since the ZEUS-2 and SEPIA observations were conducted using a single-dish, they should not be susceptible to such issues. It is also possible that the larger bandwidth of the SPIRE spectrum, compared to the ZEUS-2 or ALMA spectra, makes it sensitive to flux from broader spectral features (e.g., outflows), including a broader component of the [CII] line to which our ALMA and ZEUS-2 observations are not sensitive \citep[e.g.,][]{Maiolino2005, Maiolino2012}. We cannot fully explain the difference between the larger SPIRE [C\,{\sc ii}] flux measurement and the three consistent measurements with ZEUS-2, ALMA, and SEPIA at this time.

\subsection{Dust Opacity} \label{subsec:DustOpacity}
Before continuing, it is worth noting that even at the far-IR wavelengths discussed here, dust extinction can sometimes be non-negligible. We can estimate the wavelength-dependent dust opacity by modeling the far-IR SED as a modified blackbody \citep[e.g.,][]{Blain2003}:

\begin{equation}
\frac{S_{\nu_r}}{\Omega_{source}} = \frac{1}{(1+z_s)^3} (B_{\nu_r}(T_{dust}) - B_{\nu_r}(T_{CMB}))(1-e^{-\tau_{\nu_{r}}})
\end{equation}

where $\Omega_{source}$ is the source size, $B_{\nu_r}(T)$ is the blackbody function evaluated at the rest-frame frequency $\nu_r$ and temperature T, and $\tau_{\nu} = (\frac{\nu}{\nu_{0}})^\beta$ = $(\frac{\lambda_{0}}{\lambda})^\beta$ \citep[e.g.,][]{Draine1984}. Given that gravitational lensing conserves surface brightness, e.g. $S_{\nu}$/$\Omega_{source}$, we use both the observed flux density (uncorrected for lensing) and image-plane source-size for this calculation. We estimate an image-plane source size of $\sim$ 0.7 square arcseconds for SDP.11, by applying the same 3$\sigma$ mask as was used to calculate the 158 $\micron$ rest-frame continuum flux density. Taking this source size together with the SED-modeled dust temperature of 41K \citep{Bussmann2013}, and the continuum flux at 158 $\micron$ (rest-frame), we find that $\lambda_0 \sim$ 20 $\micron$ (assuming a dust emissivity index, $\beta$, value of 1.5). This dust opacity corresponds to corrections ranging from $\sim$ 21$\%$ at 52 $\micron$, our shortest wavelength --- and hence most highly extincted --- detected spectral line, to $\sim$ 4$\%$ at 158 $\micron$. We apply these extinction corrections to all spectral lines, in all proceeding calculations, and to the luminosities calculated in Table 1. 

\begin{deluxetable*}{cccccccccc}
\tablecaption{Spectral lines observed in SDP.11 \label{tab:lines}}
\tablecolumns{10}
\tablenum{1}
\tablewidth{0pt}
\setlength\tabcolsep{2.0pt}
\tablehead{
\colhead{Line} & \colhead{[O\,{\sc iii}]} & \colhead{[O\,{\sc i}]} & \colhead{[N\,{\sc iii}]} & \colhead{[O\,{\sc iv}]} & \colhead{[S\,{\sc iii}]} & \colhead{[N\,{\sc ii}]} & \colhead{[O\,{\sc iii}]} & \colhead{[C\,{\sc ii}] (ALMA)} & \colhead{[C\,{\sc ii}] (SPIRE)} \\
\colhead{} & \colhead{51.8\,$\micron$} & \colhead{63.2\,$\micron$} & \colhead{57.3\,$\micron$} & \colhead{25.9\,$\micron$} & \colhead{33.5\,$\micron$} & \colhead{121.9\,$\micron$} & \colhead{88.4\,$\micron$} & \colhead{157.7\,$\micron$} & \colhead{157.7\,$\micron$} \\
}
\startdata
Observed Flux $\mu$\,S$\Delta$v ($10^{-18}$\,W\,m$^{-2}$) & 7.2\,$\pm$\,1.2 & 8.2\,$\pm$\,2.6 & 2.9\,$\pm$\,1.5 & \textless\,9.4 & \textless\,7.2 & \textless\,5.8 & 8.5\,$\pm$\,1.6 & 5.9\,$\pm$\,0.2\tablenotemark{*} & 12.9\,$\pm$\,1.3 \\
Estimated Dust Opacity ($\tau$) & 0.24 & 0.18 & 0.21 & 0.68 & 0.46 & 0.07 & 0.11 & 0.05 & 0.05 \\
Dust Opacity Correction ($\%$) & 21.3 & 16.3 & 18.6 & 49.3 & 37.0 & 6.4 & 10.2 & 4.4 & 4.4 \\
$\tau$-Corrected Luminosity $\mu$\,L ($10^{10}$\,$L_{\odot}$) & 5.2 & 5.6 & 2.0 & \textless\,8.1 & \textless\,5.8 & \textless\,3.6 & 5.5 & 3.6 & 7.9 \\
\enddata
\tablecomments{``$\mu$" is the gravitational lensing magnification factor and all quantities appearing with a $\mu$ are observed quantities and hence not corrected for magnification due to gravitational lensing. Intrinsic fluxes are obtained by dividing the observed fluxes by $\mu$. Upper limits displayed in the table are 3$\sigma$ limits. $\tau$-corrected luminosities include an opacity correction to the line luminosity due to dust attenuation (see text).}
\tablenotetext{*}{This uncertainty is determined using the RMS error per beam in the ALMA [C\,{\sc ii}] 158 $\micron$ moment-zero map and does not consider uncertainty due to resolving out flux with the interferometer.}
\end{deluxetable*}

\begin{deluxetable}{ccccccccccc}
\tablecaption{Radio Continuum Observations of SDP.11 \label{tab:lines}}
\tablecolumns{5}
\tablenum{2}
\tablewidth{0pt}
\setlength\tabcolsep{2.0pt}
\tablehead{
\colhead{Observed Frequency (GHz)} & \colhead{1.43} & \colhead{6.0} & \colhead{15.0} & \colhead{33.0}
}
\startdata
$\mu$\,$S_{\nu, obs}$ ($\mu$Jy) & 642$\pm$176\tablenotemark{a} & 316$\pm$26\tablenotemark{b} & 171$\pm$14\tablenotemark{b} & 110$\pm$13 \\
$\mu$\,$S_{\nu, rest}$ ($\mu$Jy) & 231$\pm$63 & 114$\pm$9 & 61$\pm$5 & 40$\pm$5 \\
Free-Free Fraction (\%) & 5.2 & 11.5 & 18.3 & 26.5 \\
\enddata
\tablecomments{The flux densities reported in this table are observed quantities, as indicated by the ``$\mu$'s", and hence not corrected for magnification due to gravitational lensing. $S_{\nu, obs}$ and $S_{\nu, rest}$ are related by conserving $\nu S_{\nu}$ under redshift. The free-free fraction  indicates the thermal contribution to the total radio continuum at each observed frequency (see text).}
\tablenotetext{a}{\cite{Becker1995}}
\tablenotetext{b}{Ferkinhoff et al. (in prep.), \cite{Ferkinhoff2017}}
\end{deluxetable}

\subsection{H\,{\sc ii} Regions} \label{subsec:HIIRegions}

The fine-structure lines that we observe in SDP.11 allow us to determine the properties of the ionized gas within this source.

\subsubsection{Gas Density and Hardness of the Radiation Field}

The level populations of the ground-state within the O$^{++}$ ion are density sensitive, such that the [O\,{\sc iii}] 52\,$\micron$/[O\,{\sc iii}] 88\,$\micron$ line ratio yields the ionized gas density in the regime from n$_{e}$ $\sim$ 100 to 30,000 cm$^{-3}$. We find a line ratio of $\sim$ 0.9, which indicates H {\sc ii} regions in the low-density limit (n$_{e}$ $\lesssim$ 100\,cm$^{-3}$), where here we use the collision strengths from \cite{Palay2012}.

Similarly, we use the [N\,{\sc iii}] 57\,$\micron$/[N\,{\sc ii}] 122\,$\micron$ line ratio to constrain the hardness of the stellar radiation field. Given our upper limit on the [N\,{\sc ii}] 122\,$\micron$ line, we calculate a ratio of $>$ 0.6. Using the models of \cite{Rubin1985}, and the density determined from the [O\,{\sc iii}] lines, we find that this ratio is consistent with H\,{\sc ii} regions powered by stars with effective temperatures $>$ 31,000 K.

Since the radiation fields on galactic scales in star-formation-dominated galaxies are dominated by the most massive stars on the main sequence, this stellar effective temperature suggests that the starburst in SDP.11 is powered by stars of type B0 or hotter \citep{Vacca1996}, which in turn suggests that the time since the last starburst is $\lesssim$ 8 Myr \citep{Meynet2003}, or perhaps that it is still ongoing.

\subsubsection{Ionized Gas Mass}

Following \cite{Ferkinhoff2010}, we can estimate the minimum ionized gas mass required to produce the observed [O\,{\sc iii}] line flux:

\begin{equation}
M_{\rm{min}}(H^{+}) = \frac{F_{ul} 4 \pi D_{L}^{2} m_H}{\frac{g_l}{g_t} A_{ul} h \nu_{ul} X_{O^{++}}} \rm{,}
\end{equation}

\noindent where $F_{ul}$ is the flux in the fine-structure line between the upper ($u$) state and the lower ($l$) state, $D_L$ is the luminosity distance (13.65\,Gpc), $m_H$ is the mass of the hydrogen atom, $X_{O^{++}}$ is the relative abundance of O$^{++}$/H$^{+}$ within the H\,{\sc ii} regions, $g_u$ and $g_l$ are the statistical weights of the upper and lower states, respectively, $g_t$ is the partition function (the sum of the statistical weights of all relevant states available to the O$^{++}$ ion at T $=$ 8,000 K), $A_{ul}$ is the Einstein coefficient for the relevant transition \citep[2.6\,$\times$\,10$^{-5}$\,s$^{-1}$ for the {[}O\,{\sc iii}{]} 88\,$\micron$ line,][]{Wiese1966}, and $\nu_{ul}$ the frequency of that transition. Using the opacity-corrected [O\,{\sc iii}] 88\,$\micron$ line luminosity, and assuming $X_{O^{++}}$ = 5.9\,$\times$\,10$^{-4}$ \citep{Savage1996}, we obtain a minimum ionized gas mass of 1.5\,$\times$\,10$^9$ $M_{\odot}$.

\subsubsection{Gas Phase Metallicity}

Comparing the fine-structure line emission to the strength of the thermal free-free emission allows us to determine the absolute gas-phase abundance of the ions that we observe in fine-structure line emission (e.g., [O$^{++}$/H] and [N$^{++}$/H]). This is because the collisionally-excited fine-structure line emissivities scale with the product of electron and ion number density, $\epsilon_{fs} \propto n_e n_i$, while thermal free-free emissivity scales with the square of the electron number density, $\epsilon_{ff} \propto n_e n_e$, and hence the ratio of fine-structure line flux to radio free-free flux determines the absolute abundance of the relevant ion, F$_{fs}$/S$_{ff}$ $\propto n_i$/$n_e$.

To leverage this gas-phase abundance diagnostic in a number of high-redshift galaxies, we are conducting the ZEUS INvestigated Galaxy Reference Sample (ZINGRS) radio survey \citep[Ferkinhoff et al. (in prep.),][]{Ferkinhoff2017} with NSF's Karl G. Jansky Very Large Array (VLA)\footnote{The National Radio Astronomy Observatory is a facility of the National Science Foundation operated under cooperative agreement by Associated Universities, Inc.}, which aims to observe the radio continuum in high-redshift galaxies detected in FIR fine-structure lines at both 6 and 15 GHz (observed-frame), effectively measuring the strength of their free-free emission. Here we use the 6 and 15 GHz radio fluxes from the ZINGRS radio survey (Project IDs: 16B-331 and 16A-375, respectively), combined with 1.43 GHz (observed-frame) continuum from the VLA Faint Images of the Radio Sky at Twenty centimeters (FIRST) survey \citep{Becker1995} and archival VLA data taken at 33 GHz (observed-frame; PI: T. Greve, Project ID: 15B-266), to determine the contribution of the thermal free-free emission to the total radio continuum in SDP.11. These flux densities are presented in Table 2. 

We decompose the radio continuum into thermal and non-thermal components using an equation of the following form \citep[e.g.,][]{Condon1992, Klein2018}:

\begin{equation}
S_{total, r} = S_{th, 0, r} \bigg( \frac{\nu}{\nu_{0,r}} \bigg) ^{-0.1} + S_{nth, 0, r} \bigg( \frac{\nu}{\nu_{0,r}} \bigg) ^{-\alpha_{nth}}
\end{equation}

where $S_{th,0,r}$ and $S_{nth,0,r}$ are the (rest-frame) contributions to the total radio flux from the thermal and non-thermal components, respectively, at $\nu_{0,r}$ (rest-frame), and $\alpha_{nth}$ is the non-thermal power-law index.
Adopting a $\nu_{0,r}$ value of 1 GHz, as in \cite{Klein2018}, and holding $\alpha_{nth}$ fixed at 0.7, a median value for cosmic ray electrons accelerated in shocks \citep[e.g.,][]{ShuBook}, we fit for $S_{th,0,r}$ and $S_{nth,0,r}$. We find best-fit values of $S_{th,0,r}$ = 17 $\pm$ 8 $\mu$Jy and $S_{nth,0,r}$ = 675 $\pm$ 68 $\mu$Jy (see Figure 4). At a rest-frame frequency of 3.98 GHz (1.43 GHz observed-frame), where we set our calculations, the free-free contribution to the radio emission is 15 $\pm$ 7 $\mu$Jy (rest-frame). See Table 2 for the thermal contribution to the total radio SED in SDP.11 at each observed frequency.

\begin{figure}
\includegraphics[width=0.48\textwidth]{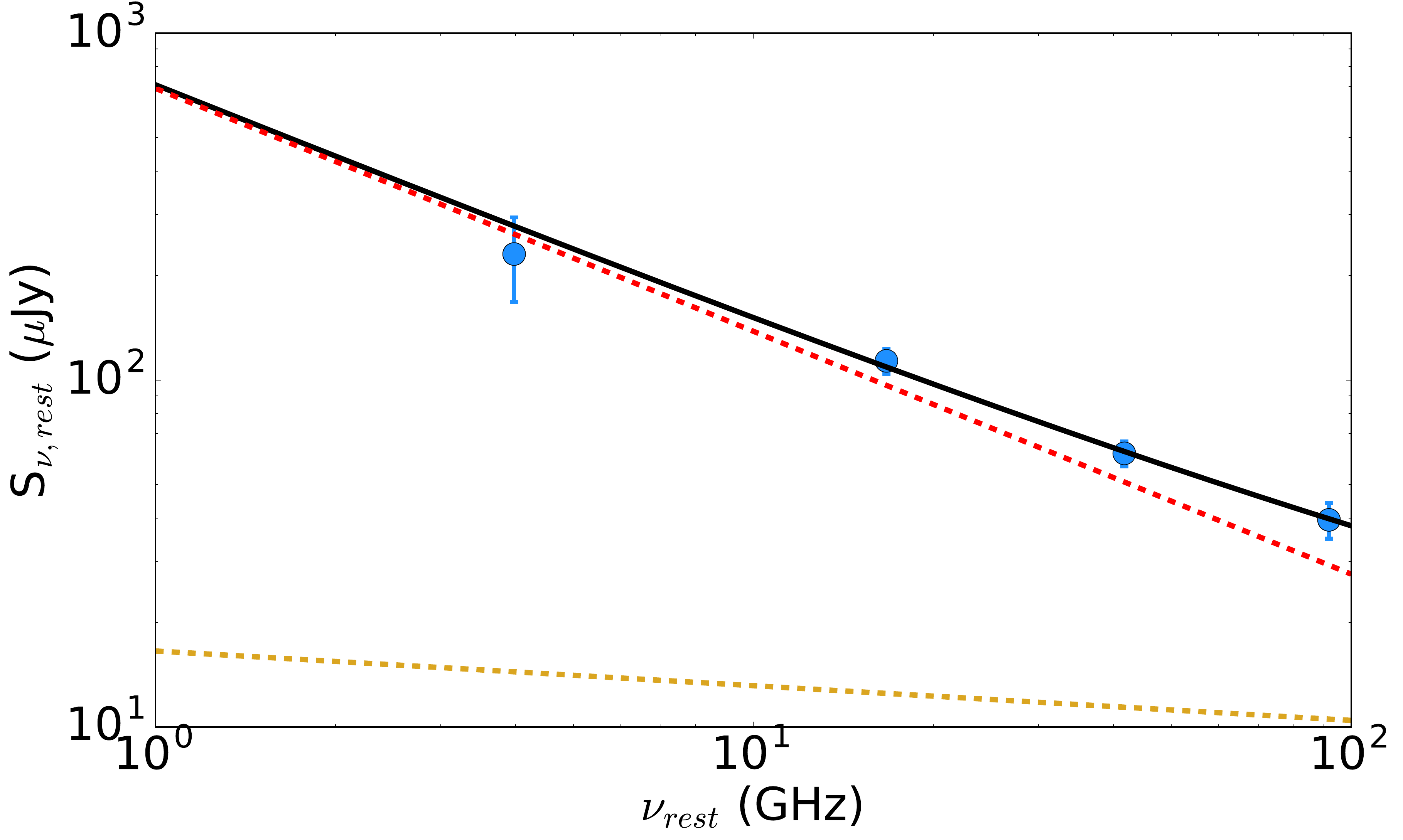}
\caption{A plot showing the radio-SED decomposition for SDP.11. The blue points represent observed continuum flux values, with associated errors, while the black solid line is the best-fit model, composed of both thermal (gold dashed) and non-thermal (red dashed) components. At a rest-frame frequency of 3.98 GHz, the free-free contribution to the total radio emission is $\sim$ 5\%.\label{fig:radio_decomposition}}
\end{figure}

The method for calculating gas-phase absolute ionic abundances from radio free-free and fine-structure line emission is well established in the literature \citep[c.f.,][]{Herter1981, Rudolph1997}, and can provide an unbiased abundance diagnostic for heavily dust obscured galaxies, especially in the early Universe where the traditional optical diagnostics can be difficult to observe and/or interpret due to dust extinction:

\begin{equation}
\frac{N_{X^i}}{N_{H^+}} = \frac{F_\lambda}{S_{\nu,r}} \frac{3.485 \times 10^{-16} T_{4}^{-0.35} \nu_{5}^{-0.1}}{\epsilon_\lambda} \left( \frac{N_e}{N_p} \right)
\end{equation}

Here $N_{X^i}$/$N_{H^+}$ is the abundance of ion i, relative to hydrogen, $F_\lambda$ is the fine-structure line flux in units of ergs s$^{-1}$ cm$^{-2}$, S$_{\nu,r}$ is the rest-frame radio free-free flux at rest-frequency $\nu$ in units of Jy, T$_4$ is the electron temperature in units of 10$^4$ K, $\nu_5$ is the radio emission frequency in units of 5 GHz, $\epsilon_\lambda$ is the emissivity per unit volume of the fine-structure line at wavelength $\lambda$, and N$_e$/N$_p$ is the electron to proton number density ratio, which accounts for the contribution of electrons from non-hydrogen atoms present in the H\,{\sc ii} regions. We use the collisional rate coefficients of \cite{Palay2012} for the [O\,{\sc iii}] lines and \cite{Stafford1994} for the [N\,{\sc iii}] line to calculate the corresponding emissivity values. We also assume N$_e$/N$_p$ = 1.05, which accounts for the electrons contributed from helium, the second most abundant element, and T$_4$ = 1, a typical value for H\,{\sc ii} regions.

Using the measured [O\,{\sc iii}] 52\,$\micron$ line flux, and free-free flux density at 3.98 GHz, together with Equation 4, we calculate [O$^{++}$/H] = 2.5 $\times$ 10$^{-4}$. Similarly, using the [N\,{\sc iii}] 57\,$\micron$ line flux, we obtain [N$^{++}$/H] = 4.9 $\times$ 10$^{-5}$. In addition to these numbers, an estimate of the fraction of O in the O$^{++}$ state, and N in the N$^{++}$ state, is required to scale back to the absolute abundances of oxygen and nitrogen.

In order to determine the [N$^{++}$/N] and [O$^{++}$/O] ratios, and hence scale our ionic abundance to total elemental abundances, an estimate for the hardness of the ambient radiation field within SDP.11 is required. While we do not have tight constraints on the hardness of the radiation field with our current observations, we can still make a reasonable estimate of the [N/O] abundance ratio in SDP.11. This is because the O$^{++}$ and N$^{++}$ ions have similar formation potentials (35.12 and 29.60 eV, respectively), such that the [N$^{++}$/N]/[O$^{++}$/O] ratio is nearly independent of stellar effective temperature for a range of parameter space. The models of \cite{Rubin1985} show that the [N$^{++}$/N]/[O$^{++}$/O] ratio is nearly constant at a value of $\sim$ 1.4 (within $\sim$ 50\%) for T$_{eff}$ $\gtrsim$ 33,000K. Adopting this value and taking the ratio of [N$^{++}$/H]/[O$^{++}$/H], calculated above, we estimate an [N/O] ratio of $\sim$ 0.14 (again within $\sim$ 50\%) in SDP.11, where solar is 0.138 \citep{Asplund2009}.

Similarly, \cite{Nagao2011} find that the [O\,{\sc iii}] (52 $\micron$ + 88 $\micron$) / [N\,{\sc iii}] 57 $\micron$ ratio scales with gas-phase metallicity, and is nearly independent of both gas density and hardness of the ambient radiation field. Our measured value of 5.4, nearly identical to their M82 value, corresponds to a gas phase metallicity of $\sim$ 0.5 - 0.7 Z$_{\odot}$, consistent with our above estimate.

\subsubsection{Star Formation Rate}

The star formation rate in SDP.11 --- calculated by converting the FIR luminosity \citep{Bussmann2013} to an IR luminosity assuming a bolometric conversion factor of 1.91 \citep{Dale2001} and then applying the scaling of \cite{Kennicutt1998} --- is $\sim$ 11,400 $\mu M_{\odot} yr^{-1}$ (uncorrected for lensing magnification). An independent estimate of the star formation rate is obtained from the free-free radio continuum. From \cite{Murphy2011}:

\begin{equation}
\begin{split}
\frac{SFR_{\nu,r}^{T}}{M_{\odot} \, yr^{-1}} = 4.6 \times 10^{-28} \\
\bigg( \frac{T_e}{10^4 K} \bigg) ^{-0.45} \bigg( \frac{\nu_r}{GHz} \bigg) ^{0.1} \bigg( \frac{L_{\nu,r}^{T}}{erg \, s^{-1} \, Hz^{-1}} \bigg) 
\end{split}
\end{equation}

where T$_e$ is the electron temperature, $\nu_r$ is the rest-frame frequency, and $L_{\nu,r}^{T}$ is the rest-frame thermal free-free luminosity at frequency $\nu$. Calculating the star formation rate using the observations at 1.43 GHz, rest-frame 3.98 GHz, we obtain $\sim$ 1,700 $\mu M_{\odot} yr^{-1}$ (uncorrected for lensing magnification). This value is a lower limit on the star formation rate in that it is based on the conversion of a star formation rate to the number of hydrogen-ionizing photons emitted by that stellar population, and then from the number of emitted hydrogen-ionizing photons to the observed free-free emission. Working backwards, the conversion from observed radio free-free emission to hydrogen-ionizing photons should be unaffected by extinction, by dust for example, however the conversion from hydrogen-ionizing photons emitted by the stellar population to those which actually ionize hydrogen atoms can be affected by dust extinction, possibly contributing to the observed discrepancy. Additionally, the differences in the calculated star formation rates could indicate the presence of an AGN, which boosts the IR luminosity and artificially inflates the star formation rate derived from that quantity, while leaving the SFR calculated from the radio free-free emission, which is disentangled from the non-thermal AGN contribution, unaffected. Recently, X-Ray observations conducted with the Chandra X-Ray Observatory \citep{Massardi2018}, and dense-gas tracers observed with ALMA \citep{Oteo2017}, have both been detected to be co-spatial with the peak of the dust continuum, suggesting the presence of an AGN within SDP.11, making this scenario plausible.

\subsection{PDRs} \label{subsec:PDRs}

\subsubsection{[C\,{\sc ii}]/FIR Ratio}

Photoelectric heating within PDRs is sensitive to the ratio of interstellar FUV radiation field strength and density (G$_0$/n) \citep[e.g.,][]{Wolfire1990}. The L$_{[C\,{\textrm{\sc ii}}]}$/L$_{FIR}$ ratio is known to trace star formation intensity, since it is sensitive to G$_0$ \citep[e.g.,][]{Stacey1991, Hailey-Dunsheath2010, Stacey2010}. This means that our spatially-resolved ALMA [C\,{\sc ii}] 158$\micron$, and underlying continuum, maps of SDP.11 allow us to examine the variations in starforming intensity across this source. 

Local ULIRGs have small [C\,{\sc ii}]/FIR ratios, $\lesssim$ 0.1\%, indicating very strong FUV fields (G$_0$ $\sim$ 1,000 - 10,000 Habing units) and intense star formation activity consistent with collision-induced star formation confined to regions of order a few hundred pc in size \citep[e.g.,][]{Diaz-Santos2017}.  In contrast, local star-forming galaxies, and many high-luminosity z $\sim$ 1 - 2 galaxies, have higher [C\,{\sc ii}]/FIR ratios, indicating more modest FUV field intensities (G$_0$ $\sim$ 100 - 1,000 Habing units) and star formation rates distributed over kpc scales in the high-redshift cases \citep[e.g.,][]{Stacey2010, Brisbin2015}. Using our image-plane [C\,{\sc ii}]-emitting source size of $\sim$ 1.9 square arcseconds, estimated from the ALMA map, and the average lensing magnification factor of 10.9 \citep{Bussmann2013}, we estimate an intrinsic source diameter of $\sim$ 0.5" ($\sim$ 4 kpc) for SDP.11, indicating extended star formation in this source. Using our spatially-resolved [C\,{\sc ii}] 158 $\micron$ line observations, together with the continuum around the line, we additionally investigate the spatial variability of the [C\,{\sc ii}]/FIR ratio across SDP.11.

The L$_{FIR}$ value per pixel was calculated by modeling the FIR SED of SDP.11 with a modified blackbody function, assuming a constant dust temperature of 41 K across the source, and a dust emissivity index ($\beta$) of 1.5 \citep[as was determined for SDP.11 using SPIRE photometry in][]{Bussmann2013}, such that the rest-frame 158 $\micron$ continuum flux scales directly to the FIR luminosity. Given that gravitational lenses are achromatic, the L$_{[C\,\textrm{\sc ii}]}$/L$_{FIR}$ ratio calculated pixel-by-pixel is expected to be unaffected by gravitational lensing, even though the lensing magnification factor may vary across the source. For this reason, we make no correction to this ratio for lensing effects.

We observe an L$_{[C\,\textrm{\sc ii}]}$/L$_{FIR}$ ratio of $\sim$ 0.02\% at the location of the peak of the dust continuum, which traces the most intense starbursting region. This low ratio suggests conditions similar to those observed in the local merger-driven ULIRGs. The L$_{[C\,\textrm{\sc ii}] }$/L$_{FIR}$ ratio increases as we move azimuthally along the Einstein ring to up to $\sim$ 0.28\% (similar to that observed in the Milky-Way, 0.3\%). We note that, given the sensitivity of our observations, we only detect dust continuum from $\sim$ 40\% of the region which is significantly detected in [C\,{\sc ii}] line emission. As such, the L$_{[C\,\textrm{\sc ii}]}$/L$_{FIR}$ ratio in regions undetected in our continuum map would be higher than those seen in Figure 5. Given the ULIRG-like L$_{[C\,\textrm{\sc ii}]}$/L$_{FIR}$ ratios that are present in this source, we cannot rule out the possibility of a merger in SDP.11, even though the velocity profile of the [C\,{\sc ii}] line is consistent with a single rotating galaxy, with a compact continuum-emitting region in the center surrounded by a more extended, $\sim$ 4 kpc diameter, [C\,{\sc ii}] emitting region.

\begin{figure}
\includegraphics[width=0.48\textwidth]{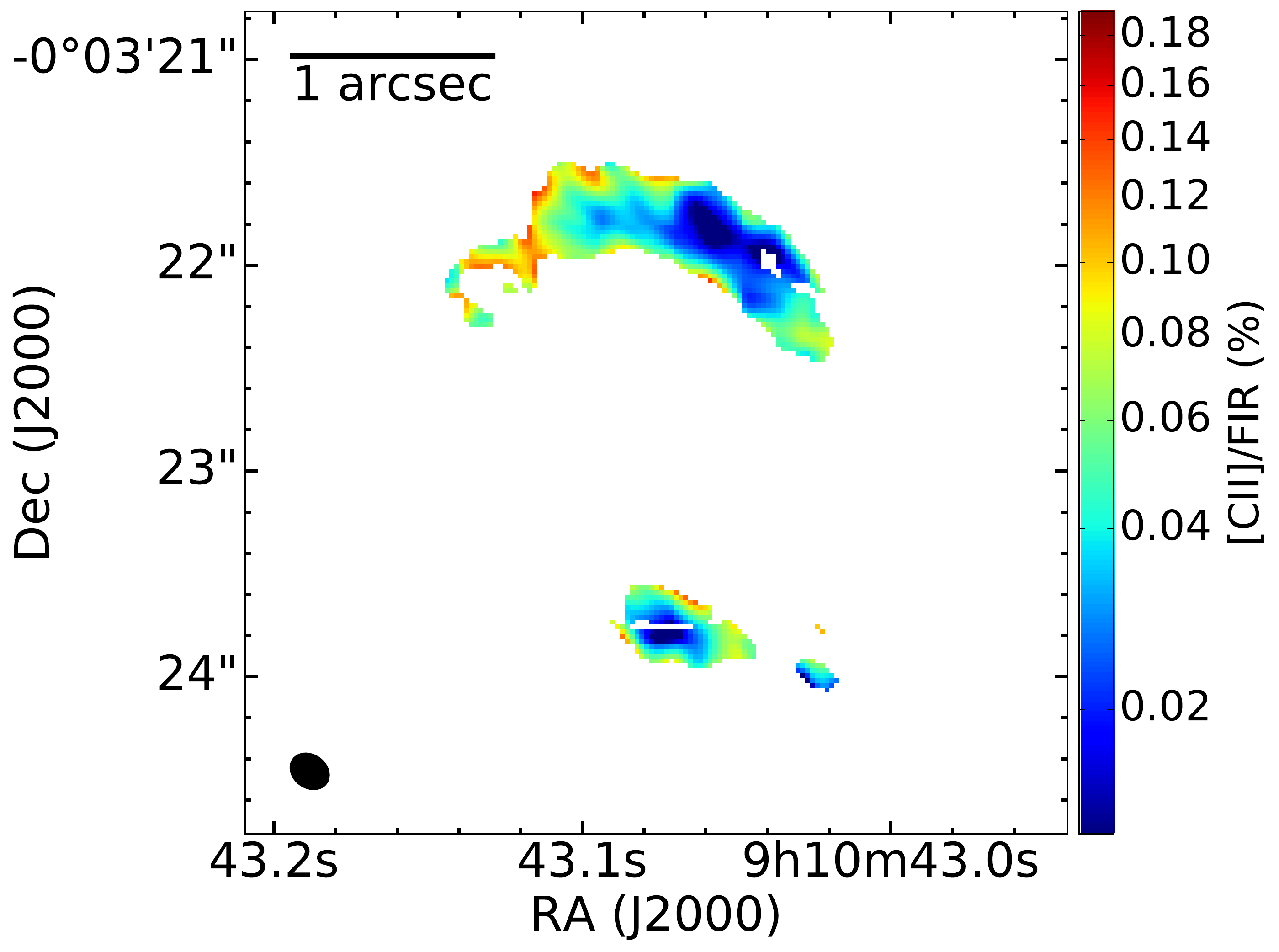}
\caption{A color map of the L$_{[C\,\textrm{\sc ii}]}$/L$_{FIR}$ ratio (plotted as a percentage) in SDP.11. The map is created by taking the ratio of the [C\,{\sc ii}] 158 $\micron$ moment-zero and L$_{FIR}$ maps, the latter created using the 158 $\micron$ (rest-frame) continuum map and assuming a constant dust temperature across the source (see text), at locations detected at a significance $>$ 3$\sigma$ in both the line and continuum maps. \label{fig:CII_FIR_ratio}}
\end{figure}

We additionally plot the L$_{[C\,\textrm{\sc ii}]}$/L$_{FIR}$ ratio vs. the star formation rate surface density ($\Sigma_{SFR}$), calculated using the L$_{FIR}$ map described above, assuming a bolometric conversion factor from L$_{FIR}$ to L$_{IR}$ of 1.91 \citep{Dale2001} and then applying the scaling of \cite{Kennicutt1998}. We then sample the map at a spatial resolution coarser than the beam to avoid correlated data points (e.g., we sample pixels from the map such that no selected pixel is within one beam width of any other selected pixel). We make no correction to the $\Sigma_{SFR}$ value calculated per pixel for gravitational lensing, which conserves surface brightness (e.g., $\Sigma_{SFR}$). Fitting a power-law to the plot of L$_{[C\,\textrm{\sc ii}]}$/L$_{FIR}$ vs. $\Sigma_{SFR}$, we find a power-law index of -0.7 (see Figure 6), indicating that the [C\,{\sc ii}] 158 $\micron$ emission increases more slowly than does L$_{FIR}$. This observed ``[C\,{\sc ii}] deficit" is in good agreement with previous studies \citep[e.g.,][see Figure 6]{Diaz-Santos2017}. This is because the [C\,{\sc ii}] line emission saturates at high UV fields, due both to the charging of grains, which reduces the efficiency of photoelectric heating of the gas, and the logarithmic growth of the C$^+$ column with UV field strength in the high-excitation limit, indicating that L$_{[C\,\textrm{\sc ii}]}$ alone is not a good measure of star formation rate. The range of values calculated for $\Sigma_{SFR}$, $\sim$ 65 - 630 M$_\odot$ yr$^{-1}$ kpc$^{-2}$, is consistent with the source-averaged value of 10$^{11.82}$ L$_\odot$ kpc$^{-2}$ ($\sim$ 218 M$_\odot$ yr$^{-1}$ kpc$^{-2}$) reported in \cite{Bussmann2013}.

\begin{figure*}
\plotone{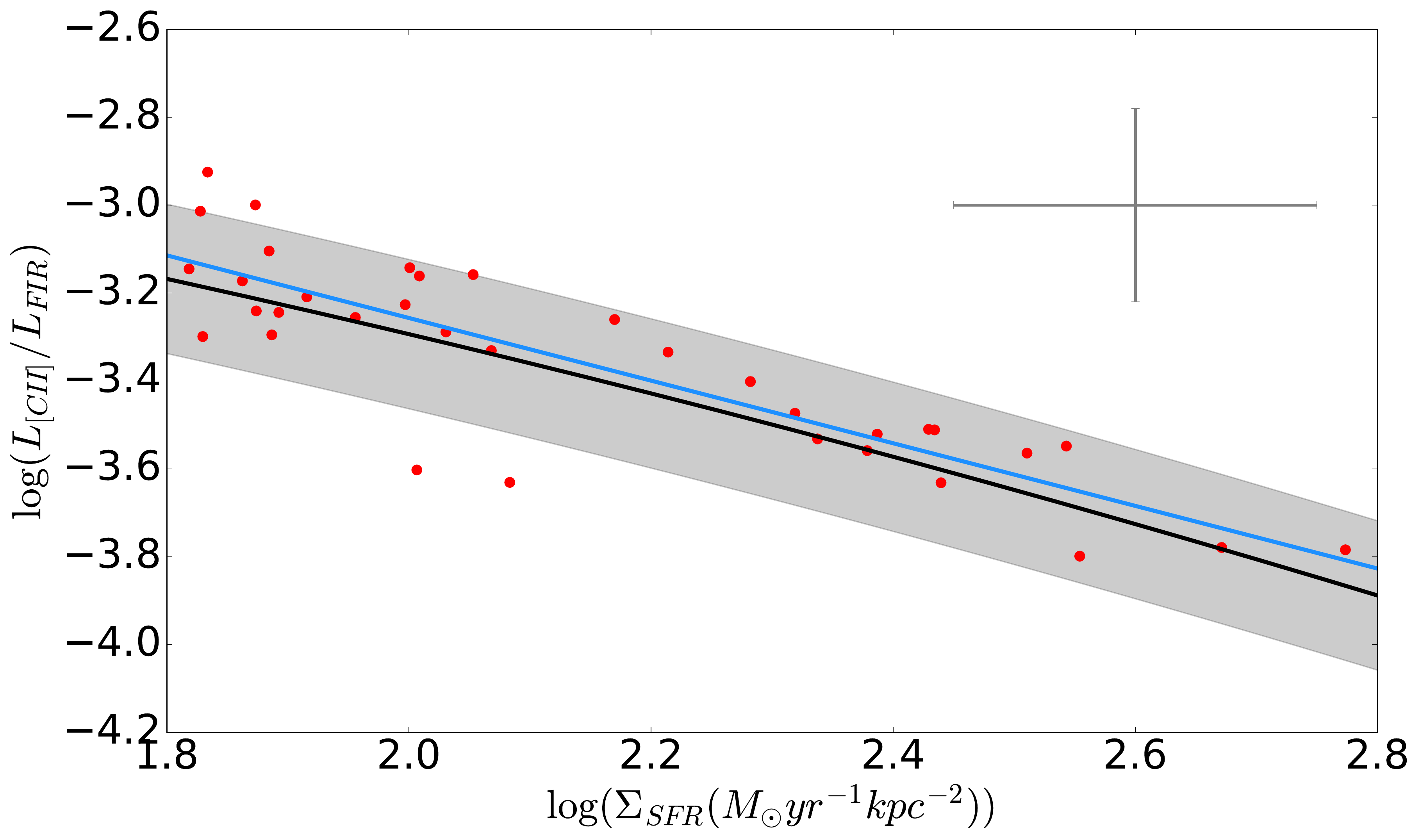}
\caption{Plot of L$_{[C\,\textrm{\sc ii}]}$/L$_{FIR}$ vs. the observed star-formation rate surface density ($\Sigma_{SFR}$) in SDP.11. The points come from sparsely sampling the 3$\sigma$-clipped ALMA [C\,{\sc ii}] 158 $\micron$ line and 158 $\micron$ (rest-frame) continuum maps at pixel separations greater than the beamsize. The best-fit line (blue) has a power-law index of -0.7, indicating that L$_{FIR}$ changes more quickly than does L$_{[C\,\textrm{\sc ii}]}$. This is because the [C\,{\sc ii}] line emission saturates at high UV fields (see text). The black line with shaded 1$\sigma$ error bounds is the best-fit from \cite{Diaz-Santos2017}, obtained for a sample of local ULIRGs. The gray error bars are representative of the worst-case error in the plot (a 3$\sigma$ detection in both the continuum and [C\,{\sc ii}] line) and are correspondingly smaller for locations detected at larger significance. \label{fig:Sigma_SFR}}
\end{figure*}

\subsubsection{PDR Mass}

We model the physical properties of the PDRs within SDP.11 using our opacity-corrected [C\,{\sc ii}] 158 $\micron$ and [O\,{\sc i}] 63 $\micron$ line luminosities, together with the FIR continuum luminosity \citep{Bussmann2013}, utilizing the PDR Toolbox \citep{Pound2008, Kaufman2006}. We obtain values of G$_0$ $\sim$ 1,800 Habing units and n $\sim$ 1,000 cm$^{-3}$, with a PDR surface temperature of $\sim$ 370 K.

Using the calculated n, G$_0$, and PDR surface temperature values in SDP.11, we can estimate the PDR mass following \cite{Hailey-Dunsheath2010}:

\begin{equation}
\begin{split}
\frac{M_{PDR}}{M_\odot} = 0.77 \bigg( \frac{0.7 L_{[C\,\textrm{\sc ii}]}}{L_\odot} \bigg) \bigg( \frac{1.4 \times 10^{-4}}{X_{C^+}} \bigg) \\
\times \frac{1 + 2 \exp(\frac{-91 K}{T}) + \frac{n_{crit}}{n}}{2 \exp(\frac{-91 K}{T})},
\end{split}
\end{equation}

where $X_{C^+}$ is the abundance of C$^+$ per hydrogen atom, taken here to be $1.4 \times 10^{-4}$ \citep{Savage1996}, n$_{crit}$ is the critical density of the [C\,{\sc ii}] 158 $\micron$ transition \citep[2,800 cm$^{-3}$,][]{Stacey2011}, and assuming that $\sim$ 70\% of the [C\,{\sc ii}] emission originates within PDRs. We calculate a PDR gas mass of $\sim$ 3.7 $\times$ 10$^{9}$ $M_{\odot}$ after correcting for the average lensing magnification factor of 10.9 \citep{Bussmann2013}.

\subsubsection{Molecular Gas}

The molecular gas in SDP.11 has been observed in several mid-J CO lines, (J$_{upper}$, J$_{lower}$) = (4-3), (5-4), (6-5), (7-6) \citep{Oteo2017, Lupu2012}. We can estimate the molecular gas mass within SDP.11 by calculating L'$_{CO(4-3)}$, using the measurements from \cite{Oteo2017}, assuming an SMG CO excitation of r$_{43/10}$ = 0.41 \citep{Bothwell2013}, and taking a ULIRG value of $\alpha_{CO}$ = 0.8 $M_{\odot}$(K km s$^{-1}$ pc$^2$)$^{-1}$ \citep[e.g.,][]{Bolatto2013}. We obtain L'$_{CO(4-3)}$ = 9.2 $\times$ 10$^{10}$ K km s$^{-1}$ pc$^2$ (uncorrected for lensing), such that the molecular gas mass is $\sim$ 1.6 $\times$ 10$^{10}$ $M_{\odot}$, after correcting for the average gravitational lensing magnification factor of 10.9 \citep{Bussmann2013}. This calculated molecular gas mass is $\sim$ 5x larger than the PDR gas mass estimated above, making the mass ratio consistent with that observed in other starburst galaxies \citep[e.g.,][]{Stacey1991}, and is $\sim$ 10x larger than the estimated ionized gas mass, consistent with ratios observed in both high-redshift \citep[e.g.,][]{Ferkinhoff2011} and nearby galaxies \citep[e.g.,][]{Lord1996, Wild1992}. With the intrinsic star formation rate of SDP.11 ($\sim$ 1,000 $M_{\odot} yr^{-1}$), this molecular gas reservoir will be depleted within $\sim$ 16 Myrs. If, instead, we assume the CO excitation of SDP.9 from \cite{Oteo2017}, which is more highly excited than is the SMG CO SLED from \cite{Bothwell2013}, the depletion timescale for the gas in SDP.11 becomes even shorter.

\subsection{Lens Modeling} \label{subsec:LensModeling}

\begin{deluxetable}{ccccccccccc}
\tablecaption{UVMCMCFIT\tablenotemark{*}-Derived Best-Fit Gravitational Lensing Parameters for SDP.11 \label{tab:gravFit}}
\tablecolumns{3}
\tablenum{3}
\tablewidth{0pt}
\tablehead{
\colhead{Parameter} & \colhead{Red Comp.} & \colhead{Blue Comp.}
}
\startdata
$\delta \rm{R.A.}_{\rm{Lens}}$ (") & 0.064 $\pm$ 0.008 & 0.064 $\pm$ 0.008\tablenotemark{\textdagger} \\
$\delta \rm{Dec.}_{\rm{Lens}}$ (") & 0.016 $\pm$ 0.008 & 0.016 $\pm$ 0.008\tablenotemark{\textdagger} \\
$\rm{Axial\,Ratio}_{\rm{Lens}}$ & 0.65 $\pm$ 0.01 & 0.65 $\pm$ 0.01\tablenotemark{\textdagger} \\
$\rm{P.A.}_{\rm{Lens}}$ (deg) & 128 $\pm$ 2 & 128 $\pm$ 2\tablenotemark{\textdagger} \\
$\rm{R}_{\rm{Einstein}}$ (") & 1.003 $\pm$ 0.004 & 1.003 $\pm$ 0.004\tablenotemark{\textdagger} \\
$\delta \rm{R.A.}_{\rm{Source}}$ (") & -0.048 $\pm$ 0.007 & -0.046 $\pm$ 0.006 \\
$\delta \rm{Dec.}_{\rm{Source}}$ (") & -0.045 $\pm$ 0.007 & 0.318 $\pm$ 0.005 \\
$\rm{Axial\,Ratio}_{\rm{Source}}$ & 0.60 $\pm$ 0.02 & 0.48 $\pm$ 0.02 \\
$\rm{P.A.}_{\rm{Source}}$ (deg) & 62 $\pm$ 3 & 74 $\pm$ 2 \\
$\rm{R}_{\rm{Eff, Source}}$ (") & 0.176 $\pm$ 0.003 & 0.166 $\pm$ 0.005 \\
$\mu$ & 11.5 $\pm$ 0.2 & 6.2 $\pm$ 0.1 \\
\enddata
\tablecomments{The red comp. and blue comp. column headings refer to the red and blue velocity components of the [C\,{\sc ii}] 158 $\micron$ line, centered at v = 155 km\,s$^{-1}$ and v = -155 km\,s$^{-1}$, respectively (where v = 0 km\,s$^{-1}$ corresponds to z = 1.7830). ``Lens" subscripts refer to properties of the foreground lensing galaxy, while ``source" subscripts refer to properties of the background, lensed, galaxy (SDP.11). The positions of the sources, $\delta R.A._{Source}$ and $\delta Dec._{Source}$, are given relative to the best-fit lens position, while the lens position, $\delta R.A._{Lens}$ and $\delta Dec._{Lens}$, is given relative to the optical centroid of the foreground lensing galaxy (9$^h$10$^m$43$^s$.07, -00$^o$03'22".91). R$_{Eff, Source}$ is the effective radius of the source in the source-plane. $\mu$ is the gravitational lensing magnification factor. See Section 3.5 for further information on the fitting procedure.}
\tablenotetext{*}{\cite{Bussmann2015}}
\tablenotetext{\textrm{\textdagger}}{The best-fit foreground lens properties obtained from the red component fit are fixed for the blue component fit, to ensure consistency (see text for further explanation).}
\end{deluxetable}

In order to recover the source-plane morphology, velocity structure, and any potential differential lensing, we perform gravitational lens modeling on the ALMA [C\,{\sc ii}] 158 $\micron$ line observations of SDP.11 using the code {\sc uvmcmcfit} \citep{Bussmann2015}. This code models the foreground lensing galaxy using a single isothermal ellipsoid (SIE) profile, which has five free parameters: the offset in both R.A. and Dec. from the user-defined center of the coordinate system, and the Einstein radius, axial ratio, and position angle of the lens. The lensed background source is modeled using a single elliptical Gaussian, and is parametrized by six free parameters: the offset in both R.A. and Dec. from the center of the user-defined coordinate system, and the intrinsic flux, axial ratio, effective radius, and position angle of the source. It then uses Markov Chain Monte Carlo (MCMC) to sample the parameter space, determining the best-fit parameters and associated uncertainties. For a more complete description of the code, see \cite{Bussmann2015}.

Before performing the gravitational lens modeling of SDP.11 using the [C\,{\sc ii}] 158 $\micron$ line, we create two moment-zero maps: one containing the red portion of the line, centered at v = 155 km\,s$^{-1}$, and the other containing the blue portion of the line, centered at -155 km\,s$^{-1}$ (where v = 0 km\,s$^{-1}$ corresponds to z = 1.7830). Each moment-zero map is collapsed over 300 km\,s$^{-1}$ in the velocity dimension. 

Given that the red component of the line is observed to be much brighter than the blue component, and hence detected at much higher significance, we perform lens modeling on that component first, requiring that the foreground lensing galaxy be located within $\pm$\,0$\farcs$2 of the centroid of the known optical source (9$^h$10$^m$43$^s$.07, -00$^o$03'22".91) obtained from Hubble imaging. We then model the blue component of the line, using the best-fit lens parameters obtained from the red-component fit (e.g., we require that both velocity components of the [C\,{\sc ii}] line are lensed by a common foreground lensing potential). See Table 3 for the gravitational lensing best-fit parameters and Figure 7 for the model images.

\begin{figure*}
\gridline{\fig{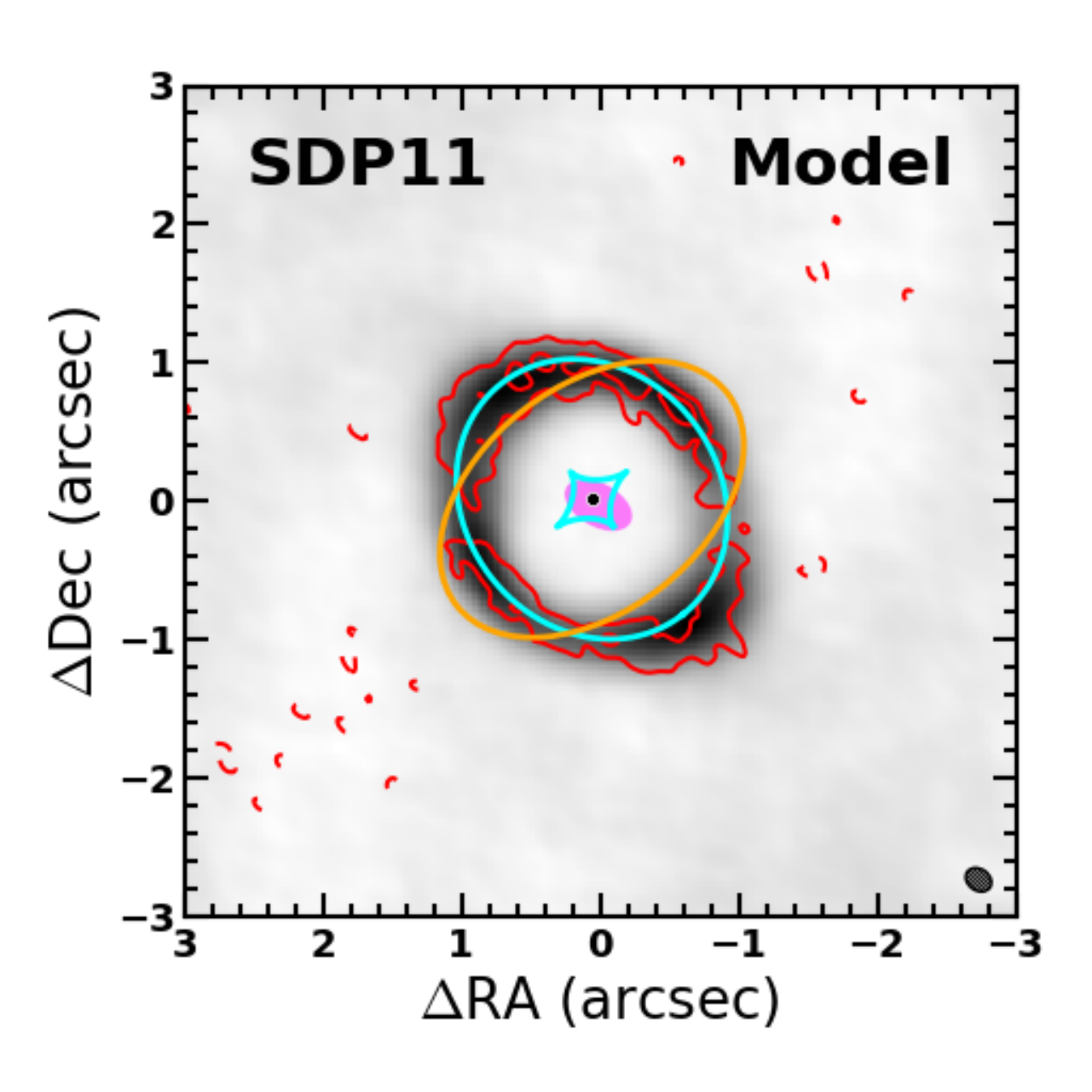}{0.5\textwidth}{(a)}
          \fig{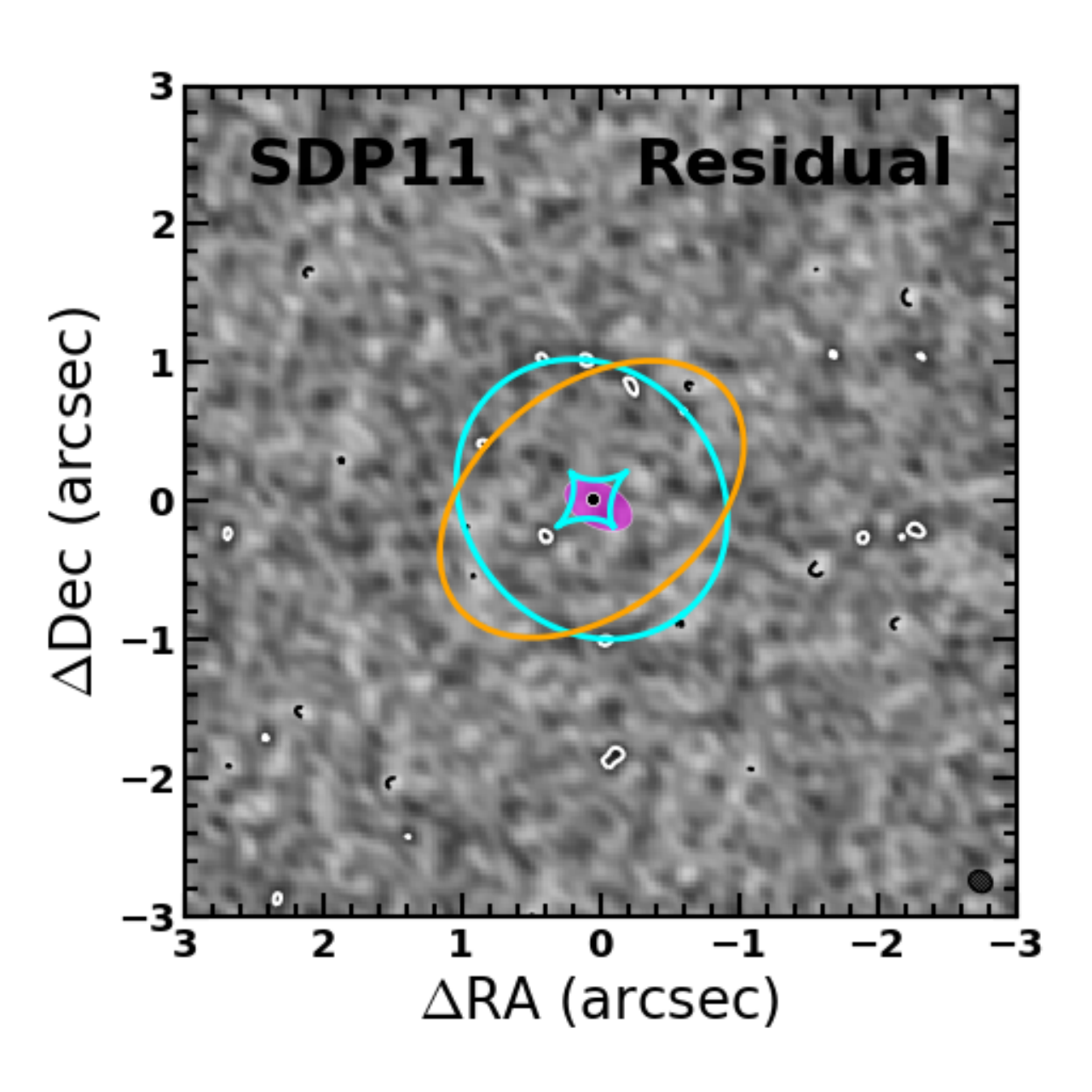}{0.5\textwidth}{(b)}
          }
\gridline{\fig{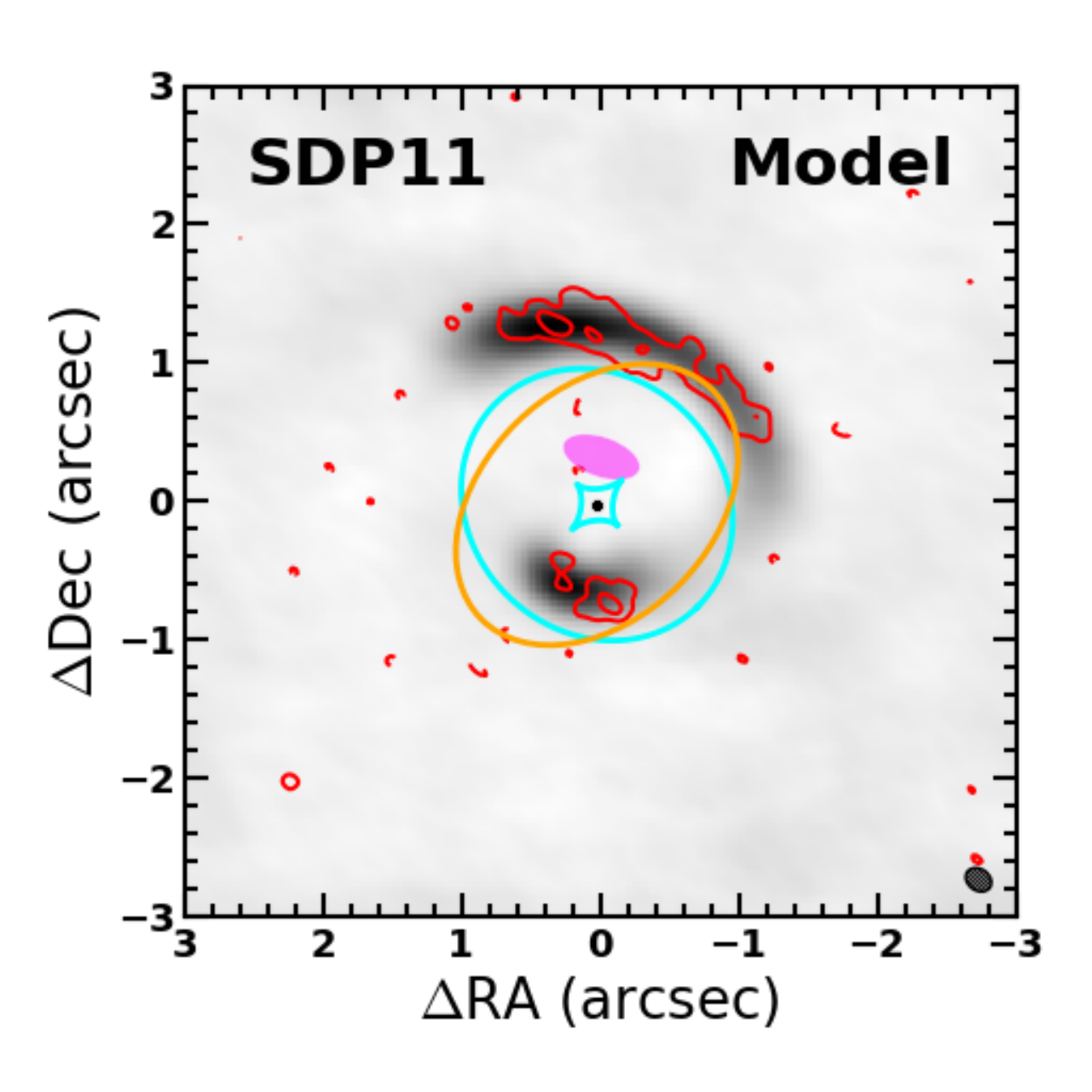}{0.5\textwidth}{(c)}
          \fig{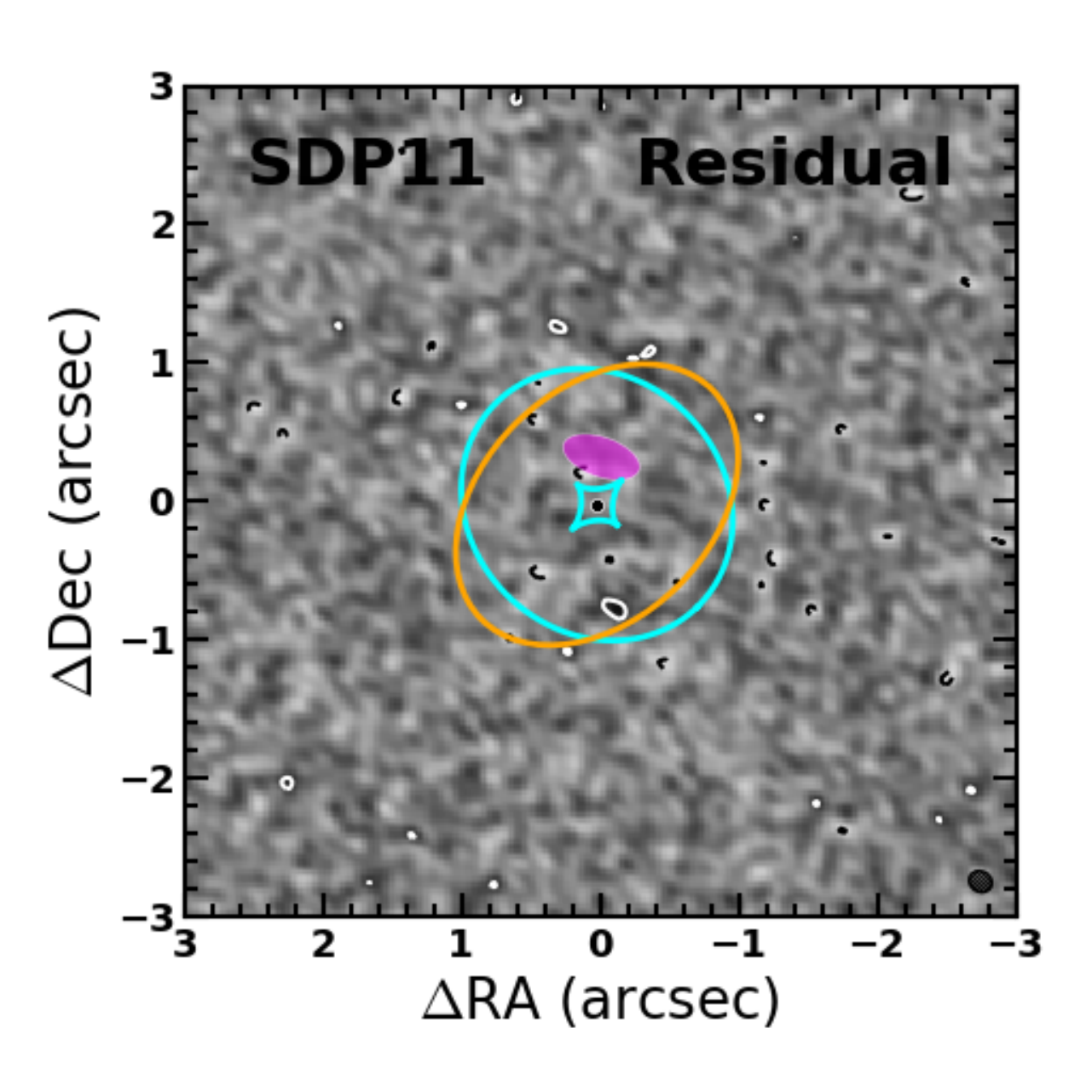}{0.5\textwidth}{(d)}
          }
\caption{(a) Gravitational lens model of the red component (v = 155 km\,s$^{-1}$) of the [C\,{\sc ii}] 158 $\micron$ line in SDP.11, created using the code {\sc uvmcmcfit} \citep{Bussmann2015}. The red contours show the [C\,{\sc ii}] 158 $\mu$m line emission, while the grayscale image is the best-fit model. The position of the foreground lensing galaxy is represented by a black dot, with its critical curve shown in orange. The half-light ellipse of the source is shown in magenta, with the caustic curve in cyan. (b) Residual map for the model shown in (a). Contours are plotted in steps of 3$\sigma$. Panels (c) and (d) are the same as (a) and (b), but for the blue component (-155 km\,s$^{-1}$) of the [C\,{\sc ii}] 158 $\mu$m line.\label{fig:gravLens}}
\end{figure*}

We find that both components of the [C\,{\sc ii}] line are well fit by a single gravitational lens located at the position of the known optical source. We further find that differential lensing is present with this lensing configuration, varying from $\mu$ = 11.5\,$\pm$\,0.2 for the red component of the line to $\mu$ = 6.2\,$\pm$\,0.1 for the blue component. After correcting for this differential lensing, the [C\,{\sc ii}] 158 $\micron$ line profile becomes much more symmetric (see Figure 8).

\cite{Dye2014} generated a pixelated reconstruction of SDP.11, based on the observed stellar emission, using an enhanced version of the semi-linear inversion method \citep[e.g.,][]{Warren2003} in the image-plane. They found that a significant external shear component, $\gamma$ = 0.23 $\pm$ 0.1, was required to describe the ellipticity of the lensed ring. This shear is attributed to a nearby edge-on spiral galaxy located $\sim$ 4$\farcs$4 to the NW of SDP.11. \cite{Dye2014} also vary the mass distribution of the foreground lensing galaxy to determine its mass density profile. We built upon the lens model presented in \cite{Bussmann2013}, based on the thermal dust emission, using the same code, UVMCMCFIT. This code uses the complete information from the visibilities sampled in the uv-plane to model the observed lensing configuration. At present, UVMCMCFIT has neither the ability to include external shear, nor the ability to vary the mass-density profile of the foreground lensing galaxy. A more detailed lens model is beyond the scope of this paper and will be the focus of future work.

If we assume that the [C\,{\sc ii}] line is indeed emitted from a rotating disk in the case of SDP.11, our gravitational lens modeling predicts that the diameter of that disk, determined from the separation of the red and blue line emitting regions in the source-plane, is $\sim$ 0$\farcs$4, or equivalently $\sim$ 3.5 kpc. This is consistent with the intrinsic source size estimated in section 3.4.1 from the [C\,{\sc ii}] image-plane source size and source-averaged gravitational lensing magnification factor.

Additionally, taking the separation between the red and blue [C\,{\sc ii}] line-emitting regions, and assuming a disk geometry, we can estimate the dynamical mass of SDP.11. For circular orbits:

\begin{equation}
M_{dyn} = \frac{v^2_{rot} r}{G},
\end{equation}

where v$_{rot}$ is the true rotational velocity of the disk, estimated from the observed velocity by correcting for the average inclination angle, $\langle$v$_{rot}$$\rangle$ $\sim$ $\frac{\pi}{2}$ v$_{obs}$ \citep[e.g.,][]{Erb2006}, r is the radius of the galaxy, and G is the gravitational constant. We estimate the radius from our gravitational lens modeling as half of the distance between the red and blue line-emitting regions (1.75 kpc). Similarly, we take half of the velocity separation between our blue and red line components, 155 km s$^{-1}$, to be the observed rotational velocity. We obtain a dynamical mass of $\sim$ 2.4 $\times$ 10$^{10}$ $M_{\odot}$. While this dynamical mass estimate depends on the assumed inclination angle of the source, taken here to be the average value, it suggests that the reported value of the stellar mass within SDP.11, $\sim$ 1.9 $\times$ 10$^{11}$ $M_{\odot}$ \citep{Negrello2014}, may be an overestimate, potentially due to contamination of the optical light by an AGN.

\begin{figure}
\includegraphics[width=0.48\textwidth]{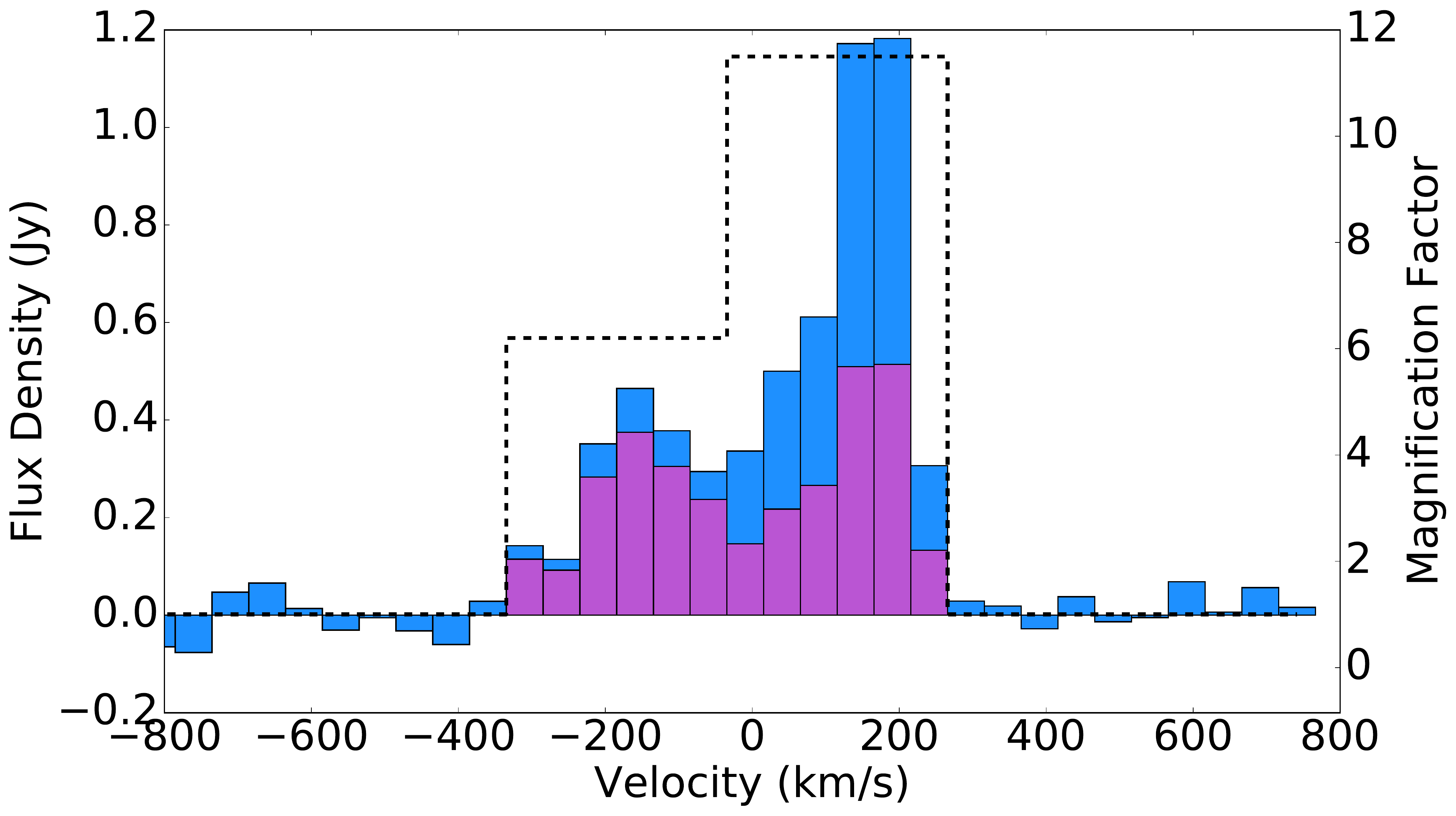}
\caption{Lens-corrected, source-integrated, ALMA [C\,{\sc ii}] 158 $\micron$ spectrum of SDP.11 (purple), scaled up by a factor of 5x for clarity. The lens-corrected spectrum is obtained by dividing the observed spectrum (blue) by the best-fit magnification factors (black dashed line) obtained using the code {\sc uvmcmcfit} \citep{Bussmann2015}. Differential lensing is present for the lensing configuration seen here, with the magnification factor varying from $\sim$ 6.2 - 11.5 across the source. After correcting for differential lensing, a more symmetric line profile is recovered. \label{fig:diffLensing}}
\end{figure}

\section{Conclusions} \label{sec:conclusions}

We have mapped the [C\,{\sc ii}] 158 $\micron$ line in SDP.11, a strongly-lensed galaxy at redshift 1.7830, at 0$\farcs$2 resolution (source-plane $\sim$ 500\,pc), using ALMA. At this resolution, the image of the gravitationally-lensed source is resolved into two spatially- and velocity-offset Einstein rings on the sky. This dataset showcases the ability of ALMA to perform high-frequency observations of high-redshift galaxies, and is one of only a handful of such results available in the literature. We have also presented detections of the [O\,{\sc iii}] 52 $\micron$, [O\,{\sc i}] 63 $\micron$, and [N\,{\sc iii}] 57 $\micron$ lines observed with Herschel/PACS.

Using the ionized gas lines from Herschel, we have modeled the H\,{\sc ii} regions of SDP.11, finding that they are heated by a starburst headed by stars hotter than spectral type B0. This stellar population constrains the age of the starburst to be $\lesssim$ 8 Myr.

Combined with multi-band radio continuum measurements, which allow us to disentangle the free-free and non-thermal contributions to the radio SED, the ionized gas lines that we detect with Herschel have allowed us to estimate the gas-phase metallicity within SDP.11. We find that the [N/O] abundance ratio in SDP.11 is consistent with solar metallicity.

Examining the [C\,{\sc ii}]/FIR ratio map of SDP.11, we find that the mean value is consistent with that of local ULIRGs, suggesting an intense starburst. We further find that the variation in the L$_{[C\,\textrm{\sc ii}]}$/L$_{FIR}$ ratio across SDP.11, when plotted against $\Sigma_{SFR}$, is best-fit with a power-law of index -0.7, indicating that the [C\,{\sc ii}] 158 $\micron$ emission increases more slowly than does L$_{FIR}$, leading to the observed ``[C\,{\sc ii}] deficit."

We have modeled the gravitational lensing configuration present for SDP.11 using the code {\sc uvmcmcfit}, finding that the position of the lens is co-spatial with a known optical source, in agreement with previous lens modeling. We further find differential lensing across SDP.11, with the lensing magnification factor varying from $\sim$ 6.2 - 11.5 across the source. After correcting for the effects of differential lensing, a more symmetric profile is recovered for the [C\,{\sc ii}] line, indicating that the starburst present here need not be the result of a major merger, with a compact starbursting region located at the center, as is the case for local ULIRGs, but may instead be star formation extended across a $\sim$ 3.5 kpc rotating disk. A more detailed study of the gravitational lensing present for this source will be the subject of a future paper.

We have estimated the ionized, PDR, and molecular gas masses in SDP.11, finding that the proportions are consistent with those of other starburst galaxies. We have additionally estimated the dynamical mass of SDP.11, finding that the previously reported stellar mass may be an overestimate, potentially due to contamination of the optical emission, attributed previously to stellar light, from an AGN.

We have recently been awarded ALMA time to map the [C\,{\sc i}] 609 and 370 $\micron$ lines, as well as the CO(4-3) and CO(7-6) lines, within SDP.11, at comparable spatial resolution to the [C\,{\sc ii}] map presented here, to further study the molecular and neutral gas within this source. Specifically, these observation will yield the spatially-resolved gas temperature, [C\,{\sc i}] 370/609 $\micron$, and CO excitation, CO(7-6)/CO(4-3), across the source, allowing for PDR modeling on sub-kiloparsec scales of SDP.11 at redshift $\sim$ 1.8.

\section*{Acknowledgments}

We thank the anonymous referee for the insightful comments and suggestions which helped to improve this manuscript. We additionally thank T. K. Daisy Leung for help in setting up the UVMCMCFIT code. C.L. acknowledges support from an NRAO Student Support Award, SOSPA3-011, and from NASA grant NNX17AF37G. D. B. acknowledges support from FONDECYT postdoctorado project 3170974.

This paper makes use of the following ALMA data: ADS/JAO.ALMA$\#$2015.1.01362.S. ALMA is a partnership of ESO (representing its member states), NSF (USA) and NINS (Japan), together with NRC (Canada) and NSC and ASIAA (Taiwan) and KASI (Republic of Korea), in cooperation with the Republic of Chile. The Joint ALMA Observatory is operated by ESO, AUI/NRAO and NAOJ.

The National Radio Astronomy Observatory is a facility of the National Science Foundation operated under cooperative agreement by Associated Universities, Inc.

\bibliography{SDP11_updated_bib.bib}

\end{document}